\pdfoutput=1
\documentclass[a4paper,12pt,english]{article}

 \usepackage[bindingoffset=0.3cm,textheight=22.5cm,hdivide={2.7cm,*,2.7cm}, vdivide={*,22cm,*}]{geometry}
 \usepackage{cite}
 \usepackage{youngtab}
 \usepackage{amsmath,amsfonts,amssymb,babel,slashed,color,empheq,tensor,amsthm}
 \usepackage{tensor}
 \usepackage[breaklinks=true]{hyperref}
 \usepackage{hyperref}

\makeatletter

\newcommand{\del}{\partial}

\def\nn{\nonumber} 

\numberwithin{equation}{section}

\def\a{\alpha} 
\def\b{\beta}\def\g{\gamma} \def\G{\Gamma}
 \def\d{\delta} 
    \def\k{\kappa}
\def\l{\lambda} \def\L{\Lambda}  \def\m{\mu}
\def\n{\nu}    \def\r{\rho}
\def\s{\sigma}  \def\t{\tau}

\def\da{{{\dot\a}}}

\def\cA{{\cal A}}  \def\cC{{\cal C}} 
 \def\cE{{\cal E}}  
\def\cG{{\cal G}} \def\cH{{\cal H}}  
 \def\cK{{\cal K}} \def\cL{{\cal L}} 
\def\cM{{\cal M}}   
  \def\cR{{\cal R}} 
 \def\cT{{\cal T}}

\def\R{{\mathbb R}} \def\C{{\mathbb C}} \def\N{{\mathbb N}}

 \def\one{\mbox{1 \kern-.59em {\rm l}}}

\def\msu{\mathfrak{su}}
\def\mso{\mathfrak{so}}

\newcommand{\eq}[1]{(\ref{#1})}

 \allowdisplaybreaks[3]

\begin{document}



\renewcommand{\title}[1]{\vspace{10mm}\noindent{\Large{\bf#1}}\vspace{8mm}} 
\newcommand{\authors}[1]{\noindent{\large #1}\vspace{5mm}}
\newcommand{\address}[1]{{\itshape #1\vspace{2mm}}}


\begin{titlepage}
\begin{flushright}
 UWThPh 2021-1
\end{flushright}
\begin{center}

\title{ {\Large Exploring the gravity sector of emergent \\[1ex]
higher-spin gravity: effective action and a solution }  }

\vskip 3mm

\authors{Stefan Fredenhagen$^{a,b}$ and Harold C.\ Steinacker$^{\,a}$}

\makeatletter{\renewcommand*{\@makefnmark}{}
\footnotetext{E-mail: \texttt{stefan.fredenhagen@univie.ac.at}, \texttt{harold.steinacker@univie.ac.at}}\makeatother}

\vskip 3mm

 \address{ 
$^{a}${\it Faculty of Physics, University of Vienna\\
Boltzmanngasse 5, 1090 Vienna, Austria  } \\[3mm]
$^{b}${\it Erwin Schr{\"o}dinger International Institute for Mathematics and Physics,\\ University of Vienna, Boltzmanngasse 9, 1090 Vienna, Austria}}

\bigskip

\vskip 1.4cm

\textbf{Abstract}
\vskip 3mm

\begin{minipage}{14.8cm}%
\vskip 3mm

We elaborate the  description of the semi-classical
gravity sector of Yang-Mills matrix models on a covariant 
quantum FLRW background. The basic geometric structure is a frame, which arises from
the Poisson structure on an underlying $S^2$ bundle over space-time.
The equations of motion for the associated Weitzenb{\"o}ck torsion
obtained in \cite{Steinacker:2020xph} are rewritten in the form of Yang-Mills-type equations for the frame.
An effective action is found which reproduces these equations of motion,
which contains an Einstein-Hilbert term
coupled to a dilaton, an axion and a Maxwell-type term for the dynamical frame.
An explicit rotationally invariant solution is found, which describes a  
gravitational field coupled to the dilaton.

\end{minipage}

\end{center}

\end{titlepage}

\tableofcontents
%
%
\section{Introduction}

Classical gravity is well described by general relativity (GR) whose dynamics is given by
the Einstein-Hilbert action. 
However, this formulation is not well suited for quantization. Moreover,
there is no straightforward way to generalize the Einstein-Hilbert action  
to noncommutative spaces, which are expected in a quantum 
theory of gravity \cite{Doplicher:1994tu}. 
This suggests that the Einstein-Hilbert action
should only be considered as an effective action, rather than  a fundamental
starting point. 
This is indeed what happens in string theory which does lead to a similar effective gravity action, but only in critical dimensions 10 or 26. 
This motivates to consider matrix models such as the 
IIB or IKKT matrix model \cite{Ishibashi:1996xs} as fundamental starting point.
While this model is known to be related to critical IIB string theory,
it opens up the possibility to study novel ways and mechanisms to obtain 
space-time and gravity on suitable noncommutative spaces or brane solutions,
cf.~\cite{Aoki:1998vn,Steinacker:2010rh,Kim:2011cr,Kim:2012mw,Nishimura:2019qal,Chaney:2015ktw,Steinacker:2017vqw,Hatakeyama:2019jyw,Klammer:2009ku,Hanada:2005vr}.

In this paper, 
we study the non-linear dynamics of the effective gravity sector which emerges 
from the IKKT or IIB matrix model on a certain type of 3+1-dimensional
covariant quantum space 
\cite{Sperling:2019xar} (cf.~\cite{Heckman:2014xha,Steinacker:2016vgf,Hasebe:2012mz,Sperling:2018xrm}).
The effective metric arises   from a dynamical frame 
 $E^{\dot\a} = \{Z^{\dot\a},.\}$, which arises from the  basic matrices $Z^{\dot\a}$
 of the matrix model via the Poisson structure in the semi-classical limit. 
The semi-classical equations of the matrix model were recast in \cite{Steinacker:2020xph} 
as non-linear geometric equations, in terms of the torsion of the Weitzenb\"ock connection associated to this frame. 
Covariance under volume-preserving diffeomorphisms is manifest in this formalism, which originates from the gauge invariance of the matrix model.
The frame and all derived objects are in general higher-spin valued, somewhat
reminiscent of Vasiliev-type higher-spin gauge theory \cite{Vasiliev:1990en,Didenko:2014dwa}.
The higher-spin modes arise here due to  the  internal structure of the brane solution as  twisted $S^2$ bundle over space-time.
The resulting 3+1-dimensional gravity theory 
was  shown to be free of ghosts in \cite{Steinacker:2019awe}, 
and to reproduce  the Ricci-flat linearized metric perturbations of general relativity including a linearized Schwarzschild solution \cite{Steinacker:2019dii}.

In the present paper, we rewrite these geometric equations for the frame
 in a more familiar form
using the standard Levi-Civita connection. 
It turns out that the totally 
antisymmetric sector of the torsion reduces on-shell to a scalar field 
 identified as axion, while the contraction of the 
torsion determines the dilaton. We obtain covariant equations of motion for 
all these fields, including generalized Maxwell-type equations for the frame.

Moreover, we  find an effective action for these geometric quantities,
in a generalized sense:
The effective action reproduces these equations of motion,  provided the frame, the dilaton and the axion are considered as independent quantities. 
This action takes a fairly familiar form involving an Einstein-Hilbert term and 
kinetic terms for the frame, the dilaton and the axion.
This action, however, becomes 
a trivial identity once all constraints of the framework are used.
Nevertheless, the action is expected to be useful, at least as a device to recover the geometric equations 
of motion in a transparent way.

The resulting gravity theory is clearly richer than GR, notably due to 
the presence of a dilaton and axion field, which is determined by the frame.
In particular, there is no manifest local Lorentz (gauge) invariance acting on the frame. That is not a problem per se, and invariance under 
volume-preserving diffeos is manifest \cite{Steinacker:2020xph}.
The vacuum solutions are
not guaranteed to be  Ricci-flat; however, Ricci-flatness does hold at 
the linearized level \cite{Steinacker:2019dii}. 

As a first step to understand the resulting physics at the non-linear level, 
we obtain in the second part of this paper
an explicit solution of the
non-linear geometric equations for a spherically symmetric static geometry centered
at some point in space.
The resulting geometry coincides with the linearized Schwarzschild geometry at the 
linearized level (as expected), but it deviates from it at the non-linear level,
with a non-vanishing dilaton contribution.
In particular there is no horizon, and the singularity at the origin is 
mild and integrable. Therefore this solution should presumably be interpreted 
as vacuum solution without matter, and its physical significance is not 
clear at this point; however, it illustrates how gravity is extended in the present
framework. From a structural point of view, it also illustrates how the higher-spin 
contributions may cancel in the effective metric.

We expect  that the present paper should be a useful starting point for finding 
further solutions of the model, and exploring the resulting theory in more depth.
The paper is organized as follows:
after a brief summary of the required background in section \ref{sec:MM-semiclass},
the main results are the covariant equations of motion in section \ref{sec:geometry},
the effective action in section \ref{sec:eff-action}, and the new solution in sections \ref{sec:solution} and \ref{sec:reconstruct}. In addition,
the appendices contain a number of new identities and structural results.

\section{Matrix model and cosmological spacetime solution}
\label{sec:MM-semiclass}

The  model underlying the present paper is the  IKKT
or IIB matrix model \cite{Ishibashi:1996xs} with mass term,
\begin{equation}
S[Z,\Psi] = {\rm Tr}\big( [Z^{\dot\a},Z^{\dot\b}][Z_{{\dot\a}},Z_{{\dot\b}}] + 2 m^2 Z_{\dot\a} Z^{\dot\a}
\,\, + \overline\Psi \Gamma_{\dot\a}[Z^{\dot\a},\Psi] \big) \ .
\label{MM-action}
\end{equation}
We will ignore the fermionic matrices $\Psi$. 
Solutions of this model are then given 
in terms of some ``matrix configurations'' consisting of
9+1 hermitian matrices $Z^{\dot\a}\in  End(\cH)$ for ${\dot\a}=0,...,9$, which satisfy the 
equations of motion (e.o.m.)
\begin{equation}
[Z^{\dot\a}, [Z_{\dot\a},Z_{\dot\b}]] = m^2 Z_{\dot\b} \ .
\label{eom-YM}
\end{equation}
Dotted indices transform under a global $SO(9,1)$, and indices are 
raised and lowered accordingly with  $\eta^{\dot\a\dot\b}$.
Fluctuations $Z^{\dot\a} \to Z^{\dot\a} + \cA^{\dot\a}$ of this background
are then governed by a (non-commutative)
gauge theory, where the fluctuations $\cA$ are typically viewed as functions 
on a background ``brane'' $\cM^{(6)}$ defined by the $Z^{\dot\a}$.

We consider a class of solutions of  \eq{eom-YM}
which are deformations of (i.e. which have similar structure as)
the following solution found in \cite{Sperling:2019xar}
\begin{align}
&\qquad & Z^{\dot\a} &= \frac{1}{R} M^{{\dot\a} 4} \ , &\dot\a =0,...,3,  &\qquad&\qquad \nn\\
&\qquad & Z^{\dot\a} &= 0,     & \dot\a =4,...,9. &\qquad &\qquad \ .
\label{background-solution}
\end{align}
Here $M^{ab} \ \in End(\cH_n), \ a,b=0,...,5$ are 
certain unitary irreducible representations  of $\mso(4,2)$ known as doubleton series, which 
depend on a label $n\in\N$. They satisfy 
\begin{align}
  [M_{ab},M_{cd}] &= i \left(\eta_{ac}M_{bd} - \eta_{ad}M_{bc} - 
\eta_{bc}M_{ad} + \eta_{bd}M_{ac}\right) \ 
 \label{M-M-relations-noncompact}
\end{align}
with $\eta_{ab} = {\rm diag}(-1,1,1,1,1,-1)$,
along with some constraints. Due to these constraints, the background 
can be  interpreted as a 6-dimensional quantized coadjoint orbit $\cM^{(6)}$,  
with a natural correspondence of matrices with functions
\begin{align}
 End(\cH_n) \cong \cC(\cM^{(6)}) \ .
\end{align}
In particular, $Z^{\dot\a}$ and the fluctuations $\cA^{\dot\a}$  can be interpreted as (quantized) functions 
on $\cM^{(6)}$.
More precisely, $\cM^{(6)}$ turns out to be a twisted sphere
bundle\footnote{In more technical terms, $\cM^{(6)}$  is  an  
$SO(3,1)$- equivariant bundle.} 
over a cosmological Friedmann-Lema{\^{i}}tre-Robertson-Walker (FLRW) spacetime,
\begin{align}
\cM^{(6)} \stackrel{loc}{\cong} \cM^{3,1} \times S^2\ ,
\label{BG-bundle}
\end{align}
in the sense that the local stabilizer of a point
$p\in\cM^{3,1}$ acts non-trivially on the fiber $S^2$. 
This means that
non-trivial harmonics on  $S^2$   lead to 
higher-spin modes on $\cM^{3,1}$, as explained in \cite{Sperling:2019xar}
or in the introductory review  \cite{Steinacker:2019fcb}. 
The geometry will be described more explicitly below in section~\ref{sec:backgroundgeometry}.

 We will work in the semi-classical limit, where commutators $[.,.]\sim i\{.,.\}$
are replaced by Poisson brackets. Then the e.o.m.~\eq{eom-YM} reduces to
 \begin{align}
 \{Z_{\dot\a}, \hat\Theta^{\dot\a\dot\b}\} = m^2 Z^{\dot\b} \ ,
 \qquad  \hat\Theta^{\dot\a\dot\b} = -\{Z^{\dot\a},Z^{\dot\b}\} \ .
 \label{eom-YM-Theta}
\end{align}
At the linearized level, the resulting higher-spin  gauge
theory was elaborated in \cite{Sperling:2019xar,Steinacker:2019awe}. A suitable formalism to understand the gravity sector 
at the non-linear level was developed in \cite{Steinacker:2020xph}, based on the 
Weitzenb\"ock connection associated to the frame 
 \begin{align}
  E_{{\dot\a}} = \{Z_{{\dot\a}},.\} = \tensor{E}{_{\dot\a}^\mu}\del_\mu\ ,  
  \qquad  \tensor{E}{_{\dot\a}^\mu} = \{Z_{{\dot\a}},x^\mu\}\ .
  \label{frame-M}
 \end{align}
In the present paper, we will develop this description further,
and find a non-trivial solution at the non-linear level.
All indices $\dot\a$ will run from 0 to 3 henceforth.

\subsection{The background geometry}
\label{sec:backgroundgeometry}

We start by reviewing the  geometry of the above background or reference
solution; more details can be found in~\cite{Sperling:2019xar}. 
The  spacetime $\cM^{3,1}$ can be described in terms of the
Cartesian coordinate functions 
\begin{align}
 x^\mu \sim \tilde R \, M^{{\mu} 5} \ , \qquad \mu=0,..,3 \ 
\end{align}
where $\tilde R$ has dimension length.
The constraints due to the doubleton representations imply 
\begin{align}
 \eta_{\mu\nu} x^\mu x^\nu \ =: -R^2 \cosh^2(\eta)  \leq - R^2 \ ,
 \qquad R = \frac{n}{2}\tilde{R} 
 \label{radial-constraint}
\end{align}
thereby defining the cosmic time variable $\eta$. 
This describes (a double cover\footnote{The two sheets of $\cM^{3,1}$ 
describe the universe before and after the Big Bounce. We will only consider 
the late-time era, where the global structure is irrelevant.} of) the segment of $\R^{3,1}$.
The equations of motion~\eqref{eom-YM} relate $R$ to the mass parameter in the matrix model by $(mR)^{2}=3$. 
The space-like 
3-hyperboloids $H^3$ defined by $\eta=const$ will be recognized  as 
equal-time slices of a $k=-1$ FLRW space-time. The $x^\mu$ transform as a vector of 
$SO(3,1)$, which should be seen as isometry group of the space-like $H^3$.

As indicated above, $\cM^{3,1}$ is the base manifold of an $S^2$ bundle $\cM^{(6)}$ \eq{BG-bundle}.
The fiber $S^2$ is described by  4 further functions 
\begin{align}
 t^\mu \sim \frac 1{R}  M^{{\mu} 4} \ , \qquad \mu=0,..,3 
\end{align}
on $\cM^{(6)}$, which also transform 
as vector of $SO(3,1)$, subject to the constraints
\begin{align}
  t_{\mu} t^{\mu}  &=  \tilde R^{-2}\, \cosh^2(\eta) \ , \\
 t_\mu x^\mu &= 0 \ .
 \label{xt-constraints}
\end{align}
Hence the $x^\mu,t^\mu$ should be viewed as functions on $\cM^{(6)}$. 
At the reference point $x = (x^0,0,0,0)$, 
these relations imply  $t^0=0$ and $t_{i} t_i = \tilde R^{-2}\, \cosh^2(\eta)$. 
While $x\in\cM^{(3,1)}$ is  invariant under its local $SO(3)$ stabilizer, the  $t^\mu$
transform as vectors under this $SO(3)$, leading to the structure of a twisted or equivariant bundle.
Accordingly,
the space of functions 
\begin{align}
 \cC(\cM^{(6)}) = \bigoplus\limits_{s\geq 0} \cC^s
 \label{cC-spin-decomp}
\end{align}
 decomposes into higher-spin modules $\cC^s$, which can be viewed as functions on $\cM^{3,1}$ taking value in 
the spin $s$ representation of the local  $SO(3)$, given by irreducible rank $s$ polynomials in $t^\mu$.

Being a coadjoint orbit, the bundle space $\cM^{(6)}$
carries a symplectic form $\omega$ which is invariant under $SO(4,2)$.
The corresponding Poisson structure is given by 
  \begin{align}
  \{x^\mu,x^\nu\} &=  \theta^{\mu\nu} = - \tilde R^2 R^2\{t^\mu, t^\nu\}
  \label{X-X-CR}\\
   \{t^\mu,x^\nu\} &= \eta^{\mu\nu} \sinh(\eta) \ .  \label{T-X-CR}   
\end{align}
The Poisson tensor $\theta^{\mu\nu}$
satisfies the constraints 
\begin{subequations}
\label{geometry-H-M}
\begin{align}
 t_\mu \theta^{\mu\nu} &= - \sinh(\eta) x^\nu \ , \\
 x_\mu \theta^{\mu\nu} &= - \tilde R^2 R^2 \sinh(\eta) t^\nu \ , \label{x-theta-contract}\\
 \eta_{\mu\nu}\theta^{\mu\kappa} \theta^{\nu\lambda} &= R^2 \tilde R^2 \eta^{\kappa \lambda} - R^2 \tilde R^4 
t^\kappa t^\lambda + \tilde R^2 x^\kappa x^\lambda  \ .
%
\end{align}
\end{subequations}
Explicitly,
\begin{align}
 \theta^{\mu\nu} &= \frac{\tilde R^2}{\cosh^2(\eta)} 
   \Big(\sinh(\eta) (x^\mu t^\nu - x^\nu t^\mu) +  \epsilon^{\mu\nu\kappa\lambda} x_\kappa t_\lambda \Big) \ .
   \label{theta-general}
\end{align}
Note that the  solution \eq{background-solution} 
amounts to $Z^{\dot\a} = t^{\dot\a}$ in the semi-classical regime.

\paragraph{Frame and metric.}

Since fluctuations $Z^{\dot\a} + \cA^{\dot\a}$ on a given background are governed by the action \eq{MM-action},
their propagation is governed by an effective metric\footnote{There are
various geometric structures which play different roles on noncommutative or quantum spaces.
The present  effective metric (which is analogous to the open string metric on branes with $B$ field) is distinct from the induced metric on 
 noncommutative branes in target space  
(which is analogous to the closed string metric).
The latter is related to the distribution of eigenvalues of the matrices, 
cf. \cite{Kim:2011cr,Aoki:1998bq}. Finally,
(quasi-) coherent states on quantum spaces are related to yet another metric \cite{Ishiki:2015saa,Steinacker:2020nva}. } 
which can be extracted from 
the kinetic term ${\rm Tr}\big( [Z^{\dot\a},\cA][Z_{{\dot\a}},\cA]\big)$. 
In the semi-classical limit, this leads to the frame \eq{frame-M}.
For the background solution, this frame takes the form
 \begin{align}
  \bar E_{{\dot\a}} = \{\bar Z_{{\dot\a}},.\} = \tensor{\bar E}{_{\dot\a}^\mu}\del_\mu\ , \qquad
  \tensor{\bar E}{_{\dot\a}^\mu} = \{\bar Z_{\dot\a},x^\mu\} = \sinh(\eta) \delta_{\dot{\a}}^{\mu}
  \label{frame-BG}
 \end{align}
which defines the auxiliary metric 
\begin{align}
  \bar\g^{\mu\nu} &= \eta^{\dot\a\dot\b}\tensor{\bar E}{_{\dot{\a}}^{\mu}}\tensor{\bar E}{_{\dot{\b}}^{\nu}} 
  = \sinh^2(\eta) \eta^{\mu\nu}\ .
\end{align}
It turns out that the effective metric $\bar{G}$ which governs the kinetic term of all propagating modes is  
a conformal rescaling  of the auxiliary metric $\bar\g^{\mu\nu}$, given by \cite{Sperling:2019xar}
\begin{align}
  \bar G^{\mu\nu} &= \frac 1{\bar\r^{2}}\, \bar\g^{\mu\nu}\ ,
   \qquad \bar\r^{2} = \rho_M\sqrt{|\bar\g|}^{-1} 
             =  \sinh^{3}(\eta) \ .
   \label{eff-metric-G}
\end{align}
Here, $|\bar\gamma|$ is the absolute value of the determinant of $(\bar\gamma_{\mu \nu})$, and 
\begin{align}
 \r_M 
 = \sinh(\eta)
 \label{rhoM-density}
\end{align}
(in Cartesian coordinates $x^\mu$)
is the density arising from the  $SO(4,1)$-invariant volume form $\Omega = \r_M d^4 x$
on $\cM^{3,1}$, which originates from the
symplectic volume form on $\cM^{(6)}$. 
$\bar G^{\mu\nu}$ is  a (hyperbolic, $k=-1$) FLRW metric,
and can be written in terms of the cosmic scale factor $a(t)$
and the comoving time $t$ as
\begin{align}
 d s^2_{\bar G} &= \bar G_{\mu\nu} d x^\mu d x^\nu 
   = -d t^2 + a^2(t)d\Sigma^2 \ ,
   \label{eff-metric-FRW}
\end{align}
where $d\Sigma^2$ 
is the $SO(3,1)$-invariant metric on the space-like $H^3$.
The cosmic scale parameter $a(t)$ is determined as
\begin{align}
a(t)^2 &=  R^2\sinh(\eta) \cosh^2(\eta) \ \stackrel {\eta\to\infty}{\approx}  \  R^2\sinh^3(\eta) 
    \label{a-eta}\\
d t &=  R \sinh(\eta)^{\frac{3}{2}} d\eta\ .
\end{align}
Note that  $a(t)$  also sets the curvature scale of the background.
In the present paper, we will focus on the {\em asymptotic regime} $a(t)\to\infty$
i.e.\ $\eta \to\infty$, considering only perturbations of the geometry on scales
far below the cosmic scale. Then the space-like metric 
$d\Sigma^2$ on $H^3 \approx \R^3$ can be approximated by a flat
metric  near $x^i = 0$, neglecting the cosmic curvature.
We can then rewrite the Cartesian coordinates around the reference point $\xi = (\xi^0,0,0,0)$ 
in terms of local spherical coordinates 
with  radius
\begin{align}
 r^2  := & \; x_i x_j \d^{ij}  \ll x_0^2 \ , \nn\\
 \eta_{\mu\nu} x^\mu x^\nu  =& -x_0^2 + r^2\ = -R^2 \cosh^2(\eta) \ .
\end{align}
In this regime, the internal sphere and the Poisson tensor 
are characterized by \cite{Steinacker:2020xph}
\begin{align}
 |t|   &\approx \tilde R^{-1} \cosh(\eta)  \nn\\
 \theta^{0i} &\stackrel{\xi}{\approx} \tilde R^2 R\, t^i  \ \gg \
  \theta^{ij} \stackrel{\xi}{\approx} \frac {\tilde R^2 R}{\sinh(\eta)}  \epsilon^{ijk} t^k  \ \stackrel{\eta\to\infty}{\sim} const \ .
\end{align}

\section{Geometric description of the non-linear regime}
\label{sec:geometry}

We recall the geometric formalism based on (a higher-spin generalization 
of) the Weitzenb\"ock 
connection and torsion. We will focus on local perturbations of the geometry
in the asymptotic regime 
$\eta\to\infty$, where the dominant contributions arise from the 
derivations along
$\cM^{3,1}$ rather than the internal directions; 
for a more detailed discussion see \cite{Steinacker:2020xph}.

\subsection{Weitzenb\"ock connection and torsion} 
\label{sec:weitzenbock}

The fundamental degrees of freedom of the matrix model are given by  matrix
configurations $Z_{{\dot\a}}$ and the associated 
vielbein 
\begin{align}
 E_{{\dot\a}} = \{Z_{{\dot\a}},.\} \ ,
 \qquad  \tensor{E}{_{\dot\a}^\mu}  = \{Z_{{\dot\a}},y^\mu\}
\end{align}
where $y^\mu$ are any local coordinate functions on $\cM^{3,1}$.
The inverse vielbein is defined as usual
\begin{align}
   \tensor{E}{^{\dot\a}_\mu}\tensor{E}{_{\dot\b}^\mu} &= \d^{\dot\a}_{\dot\b}\ , &  
    \g^{\mu\nu} &= \eta^{{{\dot\a}}{\dot\b}} \tensor{E}{_{\dot\a}^\mu} \tensor{E}{_{\dot\b}^\nu}\ ,  \nn\\
    \tensor{E}{^{\dot\a}_\nu} \tensor{E}{_{\dot\a}^\mu} &= \d^\mu_\nu \ ,  & 
  \eta_{{{\dot\a}}{\dot\b}} &= \tensor{E}{_{\dot\a}^\mu} \tensor{E}{_{\dot\b}^\nu}  \g_{\mu\nu} \ .
\end{align}
It is  natural to 
define a Weitzenb\"ock  connection  on $\cM^{3,1}$
which respects this vielbein
\begin{align}
 0 = \nabla_{\nu} \tensor{E}{_{\dot\a}^\mu} &= \del_\nu\tensor{E}{_{\dot\a}^\mu}
 + \tensor{\Gamma}{_\nu_\r^\mu} \tensor{E}{_{\dot\a}^\r} 
 \label{weizenbock-gamma-expl}
\end{align}
cf.~\cite{aldrovandi2012teleparallel}.
This connection is automatically compatible
with the metric  $\nabla \g^{\mu\nu} = 0$.
For any vector field $V^\mu$ on $\cM^{3,1}$ (possibly with a dependence on the fiber parameter $t^{\mu}$ corresponding to higher-spin modes), 
the (Weitzenb\"ock)  covariant derivative is then
\begin{align}
 \nabla_\mu V^\nu &=  \del_\mu V^\nu + \tensor{\Gamma}{_\mu_\r^\nu} V^\r  \ .
\end{align}
This connection is flat
since the frame is parallel, $\nabla E_{\dot\b} = 0$.
However, it typically has torsion,
\begin{align}
 T[X,Y] = \nabla_X Y - \nabla_Y X - [X,Y]
\end{align}
which  can be computed as \cite{Steinacker:2020xph}
\begin{align}
\tensor{T}{_\mu_\nu^{\r}} &= \tensor{\Gamma}{_{\mu}_\nu^\r}
    - \tensor{\Gamma}{_{\nu}_\mu^\r} \nn\\
\tensor{T}{_\mu_\nu^{\dot\a}} 
 &=
\tensor{T}{_\mu_\nu^{\r}} \tensor{E}{^{\dot\a}_\rho} = 
\del_\mu \tensor{E}{^{\dot\a}_\nu} - \del_\nu \tensor{E}{^{\dot\a}_\mu} \ .
    \label{torsion-explicit}
\end{align}
The torsion satisfies a Bianchi identity,
\begin{align}
 0 =  \nabla_\s \tensor{T}{_\l_\r^\mu}
 + \nabla_\l \tensor{T}{_\r_\s^\mu}  + \nabla_\r \tensor{T}{_\s_\l^\mu }
  + \tensor{T}{_\l_\r^\nu} \tensor{T}{_\nu_\s^\mu}
   + \tensor{T}{_\r_\s^\nu} \tensor{T}{_\nu_\l^\mu} 
   + \tensor{T}{_\s_\l^\nu} \tensor{T}{_\nu_\r^\mu}
 \label{Bianchi-full}
\end{align}
which follows from the first Bianchi identity for a connection with zero curvature
\cite{penrose1984spinors}, or from the Jacobi identity in the matrix model \cite{Steinacker:2020xph}. 
Its contraction gives the identity
\begin{equation}\label{Bianchi-contracted}
\nabla_\mu \tensor{T}{_\l_\r^\mu} = 0\  .
\end{equation}
In terms of the frame-valued torsion $\tensor{T}{_\mu_\nu^{\dot\a}}$ the Bianchi identity reads
\begin{equation}
0 = \partial_{\sigma}\tensor{T}{_\mu_\nu^{\dot\a}} + \partial_{\mu}\tensor{T}{_\nu_\sigma^{\dot\a}}+\partial_{\nu}\tensor{T}{_{\sigma}_\mu^{\dot\a}}\ .
\end{equation}
Viewing $\tensor{T}{_\mu_\nu^{\dot\a}}$ as components of a two-form $T^{\dot{\alpha}}$, we see from~\eqref{torsion-explicit} that it is the exterior derivative of the vielbein, and the Bianchi identity simply states that $T^{\dot{\alpha}}$ is closed,
\begin{align}
T^{\dot{\alpha}}  =  dE^{\dot\a}  
 = \frac 12\tensor{T}{_\mu_{\nu}^{\dot\a}} dx^\mu \wedge dx^\nu  \ ,\qquad dT^{\dot{\alpha}}=0\ .
 \label{Bianchi-framelike}
\end{align}
The Levi-Civita connection $\nabla^{(\g)}$  for the  metric $\g^{\mu\nu}$
is related to the Weitzenb\"ock connection via
\begin{align}
 \tensor{\Gamma}{_\mu_\nu^\r} &= \tensor{\G}{^{(\g)}_\mu_\nu^\r} + \tensor{K}{_\mu_\nu^\r} \nn\\
 \nabla_{{\mu}} V^\nu &= \nabla^{(\g)}_{{\mu}} V^\nu + \tensor{K}{_{\mu}_\r^\nu} V^\r \ .
 \label{relation-contorsion-levi}
\end{align}
Here
\begin{align}
 \tensor{K}{_{\mu}_{\nu}^{\s}}
 &= \frac 12 (\tensor{T}{_{\mu}_{\nu}^{\s}} 
             + \tensor{T}{^{\s}_{\mu}_{\nu}} 
             - \tensor{T}{_{\nu}^{\s}_{\mu}})
 = -  \tensor{K}{_\mu^\s_\nu} 
  \label{Levi-contorsion-basic}
\end{align}
is the contorsion of the  Weitzenb\"ock connection, 
which is antisymmetric in the last 2 indices. It carries the same information as the torsion, which can be reconstructed as $\tensor{T}{_{\mu}_{\nu}^{\s}}= \tensor{K}{_{\mu}_{\nu}^{\s}}-\tensor{K}{_{\nu}_{\mu}^{\s}}$.
Note that all these quantities can in general take values in 
the higher-spin algebra $\cC$ of functions on $\cM^{(6)}$.
In particular, the torsion of 
the cosmic background  is given by (see (7.26) in \cite{Steinacker:2020xph}) 
 \begin{align}
 \tensor{\bar T}{_{\r}_{\s}^\mu} 
  \approx  \frac{1}{a(t)^2}\big( \d_\s^{\mu} \t_\r - \d^\mu_\r \t_\s \big) 
  \label{torsion-BG-explicit}
\end{align}
where $\t_\mu = \bar G_{\mu\nu}\t^\nu$ and
\begin{align}
 \t = x^\mu \del_\mu = a(t)\del_t
 \label{tau-def}
\end{align}
is the time-like vector field on the FLRW background.

\paragraph{Effective metric.}

Similarly to the discussion of the background geometry, the form of the kinetic term for fluctuations around a matrix model solution leads us to introduce an effective metric $G^{\mu\nu}$ (see~\eqref{eff-metric-G}) that differs from $\g^{\mu\nu}$ by a conformal factor 
(see eq.~(4.10) in~\cite{Steinacker:2020xph}):
\begin{align}
 G^{\mu\nu} := \frac 1{\r^2} \g^{\mu\nu}\ , \qquad 
 \r^2 &= \r_M \sqrt{|\g|}^{-1} 
 \label{eff-metric}
\end{align}
where $\r_M d^4 y$ is the symplectic volume form. We shall therefore denote the scalar field $\r$ as {\em dilaton}.
The Levi-Civita connection $\nabla^{(G)}$  for the effective metric $G^{\mu\nu}$ 
 is then
\begin{align}
  \tensor{\G}{^{(G)}_\mu_\nu^\s}
 = \tensor{\G}{_\mu_\nu^\s} + \d^\s_\nu \r^{-1} \del_\mu \r - \tensor{\cK}{_\mu_\nu^\s} \ .
\label{LC-contorsion-eff}
\end{align}
Here
\begin{align}
 \tensor{\cK}{_\mu_\nu^\s} &=
   \tensor{K}{_{\mu}_{\nu}^{\s}}
  + \Big(G_{\mu\nu}\r^{-1} \del^\s \r - \d^\s_{\mu}\r^{-1}\del_\nu \r\Big) 
   = -  \tensor{\cK}{_\mu^\s_\nu} \label{Levi-contorsion-full}\\
\tensor{\cT}{_{\mu}_{\nu}^\s} &=  \tensor{T}{_{\mu}_{\nu}^\s} + \r^{-1} \big(\d_{\nu}^\s\del_{\mu}     
  \rho - \d_{\mu}^\s\del_{\nu}\rho \big) \ 
    \label{tilde-T-T}
\end{align}
is the Weitzenb\"ock contorsion and torsion tensor
of the effective frame 
\begin{align}
  \tensor{\cE}{^{\dot\a}_\mu} = \r \tensor{E}{^{\dot\a}_\mu} \ .
  \label{eff-frame}
\end{align}
This allows us to rewrite the effective Levi-Civita connection in terms of the 
Weitzenb\"ock connection and the contorsion:
\begin{align}
 \nabla^{(G)}_\mu V^{\s} &= \nabla_\mu V^{\s}
 - \tensor{\cK}{_\mu_\nu^\s} V^\nu
  +  \r^{-1}\del_{\mu}\r\, V^\s \ .
   \label{nabla-nablaLC-rel}
\end{align}
Accordingly, its indices should be raised and lowered with $G^{\mu\nu}$.
To avoid any confusion,
it is safer to write 
all connection and (con)torsion symbols with two lower and 
one upper index, where no ambiguity arises.
Calligraphic fonts indicate the effective frame.

The Jacobi identity in the matrix model implies a relation between the trace of the torsion and the dilaton~\cite[Lemma 5.2]{Steinacker:2020xph},
\begin{align}
   \tensor{T}{_\mu_{\s}^\mu} = \tensor{K}{_\mu_{\s}^\mu} &= \frac{2}{\rho}\del_\s\r \ .
 \label{tilde-T-T-contract}
 \end{align}
For the
rescaled frame resp.\  effective metric, we analogously have
 \begin{align}
  \tensor{\cT}{_\mu_{\s}^\mu}  = \tensor{\cK}{_\mu_{\s}^\mu} 
  &= - \r^{-1} \del_\s\rho \ .
 \label{torsion-contorsion-contract}
\end{align}

\subsection{Equations of motion for the torsion and frame}
\label{sec:conservation-law}

Starting from the semi-classical e.o.m.~\eq{eom-YM-Theta}, 
 the following e.o.m.\ for the torsion in vacuum  was obtained in \cite{Steinacker:2020xph}
\begin{align}
\boxed{\ 
 \nabla_\nu \tensor{T}{^\nu_\r_\mu}  +   \tensor{T}{_\nu^{\s}_\mu}\tensor{T}{_\s_\r^\nu} 
  = m^2 \g_{\r\mu} \  .
  \ }
\label{torsion-eq-nonlin-coord}
\end{align}
The non-linear equation of motion encodes the non-linear structure of the Yang-Mills 
equations of motion \eq{eom-YM-Theta}.
It can be rewritten in terms of the Levi-Civita connection
using \eq{nabla-nablaLC-rel} and~\eqref{torsion-contorsion-contract},
\begin{align}
m^2 \g_{\r\mu}  &= \nabla^{(G)}_\nu \tensor{T}{^\nu_\r_\mu}  
   - \tensor{\cK}{_\nu_\r^\s}\tensor{T}{^\nu_\s_\mu} 
   - \tensor{\cK}{_\nu_\mu^\s}\tensor{T}{^\nu_\r_\s}
   + \tensor{T}{_\nu^{\s}_\mu}\tensor{T}{_\s_\r^\nu}  \nn\\
&=  \nabla^{(G)}_\nu \tensor{T}{^\nu_\r_\mu}  
     + \frac 12(-\tensor{T}{^{\lambda}_{\nu}_{\rho}}\tensor{T}{^{\nu}_{\lambda}_{\mu}} - \tensor{T}{_{\rho}^{\lambda}_{\nu}}\tensor{T}{^{\nu}_{\lambda}_{\mu}} -\tensor{T}{^{\nu}_{\rho}_{\lambda}}\tensor{T}{_{\nu}_{\mu}^{\lambda}} + \tensor{T}{^{\nu}_{\rho}_{\lambda}}\tensor{T}{_{\mu}^{\lambda}_{\nu}}    ) \nn\\
 &\quad -  \r^{-1}\del_\s\r (\tensor{T}{_\r^\s_\mu}  +\tensor{T}{_\mu_\r^\s})
         + 2 \r^{-2}\del_\mu\r \del_\r\r \ .
         \label{eom-torsion-LeviC}
\end{align}
Together with the Bianchi identity~\eqref{Bianchi-full}
(and the flatness of $\nabla$), this captures the dynamical content 
of the model.

\paragraph{The antisymmetric part of the e.o.m.}

When we consider the e.o.m.~\eqref{torsion-eq-nonlin-coord}, we observe that the right hand side is proportional to the metric and hence is a symmetric tensor. The left hand side is not automatically symmetric, and the e.o.m.\ can be split into two components: one that requires that the antisymmetric part of the left hand side vanishes, and the remaining symmetric part. 

The antisymmetric part of~\eqref{torsion-eq-nonlin-coord} reads
\begin{equation}\label{AS-eom-new}
 \nabla_\nu \big(\tensor{T}{^\nu_\r_\mu} - \tensor{T}{^\nu_\mu_\r} \big) 
 +  2\, \tensor{T}{_\nu^{\s}_{[\mu|}}\tensor{T}{_\s_{|\r]}^\nu} 
  = 0
\end{equation}
where the square brackets denote anti-symmetrization. When we introduce the totally antisymmetric component of the torsion as (cf.~appendix \ref{sec:AS-torsion-app})
\begin{align}
  \tensor{T}{^{(AS)}^{\nu}_{\r\mu}}  &=  \tensor{T}{^\nu_\r_\mu} + \tensor{T}{_\mu^\nu_\r} + \tensor{T}{_\r_\mu^\nu}\ ,
  \label{T-AS}
\end{align}
we can rewrite the antisymmetric e.o.m.~\eqref{AS-eom-new} as
\begin{equation}\label{AS-eom-new-2}
\nabla_\nu \tensor{T}{^{(AS)}^\nu_\r_\mu} =  \tensor{T}{^{\sigma}_{\nu}_{[\rho}} \tensor{T}{^{(AS)}_{\mu ]}_{\sigma}^{\nu}}\ ,
\end{equation}
where the contracted Bianchi identity~\eqref{Bianchi-contracted} has been used.

The antisymmetric e.o.m.\ can be written in a more convenient form using the Levi-Civita covariant derivative with respect to the effective metric $G$. Using the relation~\eqref{nabla-nablaLC-rel} between $\nabla^{(G)}$ and $\nabla$, as well as the result on the trace of the contorsion~\eqref{torsion-contorsion-contract} we find
\begin{align}
\rho^{-2} \nabla^{(G)}_{\nu} \big(\rho^{2}\tensor{T}{^{(AS)}^{\nu}_{\rho}_{\mu}} \big)
& = \nabla_{\nu} \tensor{T}{^{(AS)}^{\nu}_{\rho}_{\mu}} + 2\,\tensor{K}{_{\nu}_{[\rho |}^{\sigma}} \tensor{T}{^{(AS)}^{\nu}_{\sigma}_{|\mu ]}} \\
& =  \nabla_{\nu} \tensor{T}{^{(AS)}^{\nu}_{\rho}_{\mu}} - \tensor{T}{^{\sigma}_{\nu}_{[\rho |}}\tensor{T}{^{(AS)}_{|\mu ]}_{\sigma}^{\nu}}\ ,
\end{align}
which vanishes due to the equation of motion. Hence
\begin{align}\label{div-T-AS}
\boxed{\ 
  \r^{-2}\nabla^{(G)}_\nu (\r^2 {T^{(AS)\nu}}_{\r\mu}) = 0 \ .
  \ }
\end{align}
This means that it is consistent to set $T^{(AS)}_{\r\s\mu}=0$, which holds for the 
background solution $\bar{T}$ in~\eqref{torsion-BG-explicit}.  
We can interpret the antisymmetric part $T^{(AS)\nu}{}_{\rho \mu}$ as the components of a 3-form,
\begin{equation}
 \frac{1}{3!} G_{\nu \nu '}T^{(AS)\nu'}{}_{\rho \mu} dx^{\nu}\wedge dx^{\rho}\wedge dx^{\mu}
= \rho^{2} T^{\dot{\alpha}}\wedge E_{\dot{\alpha}} \ .
\end{equation} 
The equation of motion~\eqref{div-T-AS} of $T^{(AS)}$ can then be rewritten -- via the Hodge star with respect to $G$ -- as
\begin{equation}
*d* (\rho^{4}  T^{\dot{\alpha}}\wedge E_{\dot{\alpha}}) = 0\ .
\end{equation}
Expressing $T^{(AS)\nu}{}_{\rho \mu}$ in terms of the $*$-dual 1-form $T=T_{\sigma}dx^{\sigma}$,
\begin{align}
 \tensor{T}{^{(AS)}^\nu_\r_\mu} &=: -\sqrt{|G|}G^{\n\n'}\varepsilon_{\n'\r\mu\s} G^{\s\s'} T_{\s'} \qquad \Longleftrightarrow \qquad \rho^{2} T^{\dot{\alpha}}\wedge E_{\dot{\alpha}} = *T  \ ,
 \label{T-AS-dual-3}
\end{align}
the equation of motion~\eqref{div-T-AS} for $T^{(AS)}$ becomes 
\begin{align}
 0 &=  \r^{-2}\nabla^{(G)}_\nu (\sqrt{|G|}G^{\n\n'}\tensor{\varepsilon}{_{\n'}_\r_\mu_\s}\r^2 G^{\s\s'}T_\s) \nn\\
  &= \r^{-2}\sqrt{|G|}G^{\s\s'}G^{\n\n'}\tensor{\varepsilon}{_\nu_\r_\mu_\s}\del_{\nu'} (\r^2 T_\s) \ .
\label{eom-AS-T}
\end{align}
In terms of differential forms this relation simply reads
\begin{equation}
d (\rho^{2}T) = 0 \ .
\end{equation}
This in turn means that $T_\mu$ can be written on-shell as 
\begin{align}
 \r^2 T_\mu = \del_\mu \tilde\r 
 \label{T-del-onshell}
\end{align}
in terms of a scalar field $\tilde\rho$, which will be denoted as {\em axion}, for reasons that will 
become clear below.
Hence the anti-symmetric part of the e.o.m.\ for the torsion 
reduces the 4 dof of  $T^{(AS)}$ to the scalar field $\tilde\r$, while the
remaining 3 dof of the general frame disappear on-shell. 

The Bianchi identity~\eqref{Bianchi-framelike} for $T^{\dot{\alpha}}$ implies 
\begin{equation}
*d* (\rho^{-2}T)= *d (T^{\dot{\alpha}}\wedge E_{\dot{\alpha}}) = * (T^{\dot{\alpha}}\wedge T_{\dot{\alpha}})\ .
\end{equation}
Expressing $T_{\mu}$ on-shell in terms of $\tilde{\rho}$ via~\eqref{T-del-onshell} results in 
\begin{equation}
*d * (\rho^{-4}d\tilde{\rho}) = * (T^{\dot{\alpha}}\wedge T_{\dot{\alpha}})\ ,
\end{equation}
or, in components (for an explicit calculation see~\eqref{div-Tmu-id}),
\begin{align}
\boxed{\ 
  \nabla_{(G)}^\mu (\r^{-4}\del_\mu\tilde\r)
  = \frac 14 \sqrt{|G|}^{-1} \varepsilon^{\n\r\mu\k}
 \tensor{T}{_\nu_\r^{\dot\a}} \tensor{T}{_\k_\mu_{\dot\a}}  
 \ . \  }\ 
 \label{div-T-dEdE}
\end{align}
Thus $\tilde\r$ is recognized as an axion-like field.

\paragraph{Eom for $\r$.}

Taking the trace of the equation of motion~\eqref{torsion-eq-nonlin-coord} for $T$ yields
\begin{equation}
\gamma^{\nu \sigma}\nabla_\sigma \tensor{T}{_\nu_\r^{\rho}}  +   \gamma^{\mu \rho}\,\tensor{T}{_\nu^{\s}_\mu}\tensor{T}{_\s_\r^\nu} 
  = 4m^2 \ . 
\end{equation}
We now rewrite $\gamma$ in terms of the effective metric $G$ via~\eqref{eff-metric}, and express $\nabla$ in terms of $\nabla^{(G)}$ using~\eqref{nabla-nablaLC-rel}. Using the results~\eqref{tilde-T-T-contract} and~\eqref{torsion-contorsion-contract} for the trace of the (con-)torsion, we obtain 
\begin{equation}
-2\rho^{2}G^{\mu \nu}\nabla_{\mu}^{(G)} (\rho^{-1}\partial_{\nu}\rho) + \rho^{2}G^{\mu\rho}\,\tensor{T}{_\nu^{\s}_\mu}\tensor{T}{_\s_\r^\nu} = 4m^{2}\ .
\end{equation}
Finally, rewriting the quadratic term in the torsion using~\eqref{TAS-contract} and replacing $T_{\mu}$ 
by its on-shell value~\eqref{T-del-onshell},  we arrive at (cf.~(5.50) in \cite{Steinacker:2020xph})
\begin{align}
\boxed{\ 
 - \nabla^{\mu}_{(G)}\big(\r^{-1}\del_\mu\r)         
  = 2 \r^{-2} m^2 + \frac 14\tensor{T}{_\mu^\s_\r}\tensor{T}{_\nu_\s^\r} G^{\mu\nu}
  + \frac 1{2} \r^{-4} G^{\mu\nu} \del_\mu\tilde\r \del_\nu\tilde\r 
  \ .\  }
 \label{Box-r-onshell}
\end{align}
We have thus identified two scalar fields $\r$ and $\tilde\r$
which encode the trace and the totally anti-symmetric components of the 
torsion tensor, respectively. Both satisfy  second-order equations of 
motion sourced by the torsion.  
 This reflects the fact that in contrast to general relativity, there is no 
local Lorentz invariance for the frame. 
In the present theory, the frame is a physical object  which 
satisfies the divergence constraint 
\begin{align}
  \nabla^{(G)}_\nu(\r^{-2}\tensor{E}{_{\dot\a}^{\nu}}) = 0
 \label{frame-div-free}
\end{align} 
which is a consequence of the relation~\eqref{tilde-T-T-contract} between the contraction of the torsion and the dilaton (as shown in appendix~\ref{sec:div}), and gives rise to extra physical degrees 
of freedom encoded in $\r$ and $\tilde\r$. These have no counterpart 
in general relativity. Similar fields are known to arise in various generalizations of 
general relativity \cite{Galtsov:1995zm,Bakas:1996dz,Clement:1996nh}.

\paragraph{Equation for frame-valued torsion.}

The equation of motion \eq{torsion-eq-nonlin-coord} for the torsion can be rewritten in terms of the
frame-valued torsion 
$\tensor{T}{_\mu_\nu^{\dot\a}} = \tensor{T}{_\mu_\nu^\s}\tensor{E}{^{\dot\a}_\s}$,
\begin{align}
 \g^{\nu\nu'}\nabla_{\nu'} \tensor{T}{_\nu_\r^{\dot\a}}  +  \tensor{T}{_\s_\r^\nu} \tensor{T}{_\nu^{\s}^{\dot\a}}
   &= m^2 \tensor{E}{^{\dot\a}_\r}  \ ,
\end{align}
where one has used that the vielbein is covariantly constant.

This equation can be expressed in terms of the Levi-Civita connection with respect to the effective metric
as follows
\begin{align}
%
   %
   \g^{\nu\nu'}(\nabla^{(G)}_{\nu'} \tensor{T}{_\nu_\r^{\dot\a}} 
  - \tensor{\cK}{_{\nu'}_\r^\s} \tensor{T}{_\nu_\s^{\dot\a}}
   + \r^{-1}\del_{\nu'}\r \tensor{T}{_\nu_\r^{\dot\a}})
 +  \tensor{T}{_\s_\r^\nu} \tensor{T}{_\nu^{\s}^{\dot\a}}
   &= m^2 \tensor{E}{^{\dot\a}_\r}
\end{align}
using \eq{torsion-contorsion-contract}.
The second term can be rewritten using
\begin{align}
  - \g^{\nu\nu'}\tensor{\cK}{_{\nu'}_\r^\s} \tensor{T}{_\nu_\s^{\dot\a}}
  +  \tensor{T}{^\s_\r^\nu} \tensor{T}{_\nu_{\s}^{\dot\a}}
%
%
  &= - \frac 12 \r^2 G^{\nu\nu'}\tensor{T}{^{(AS)}_{\nu'}_\r^\s}\tensor{T}{_\nu_{\s}^{\dot\a}}  
   - \r G^{\s\s'}\del_{\s'}\r \tensor{T}{_\r_{\s}^{\dot\a}} 
\end{align}
where $G^{\nu\nu'} = \r^{-2}\g^{\nu\nu'}$, which results in
\begin{align}
 \r^{-2}\nabla_{(G)}^\nu (\r^2 \tensor{T}{_\nu_\r^{\dot\a}} ) 
  = \frac 12  G^{\nu\nu'}\tensor{T}{^{(AS)}_{\nu'}_\r^\s} \tensor{T}{_\nu_{\s}^{\dot\a}} 
   + \r^{-2} m^2 \tensor{E}{^{\dot\a}_\r} \ .
  \label{torsion-frame-eom}
\end{align}
Expressing $T^{(AS)}$ in terms of the dual 1-form $T$ (see \eq{T-AS-dual}), we obtain (using \eq{epsilon-lift})
 \begin{align}  
 \nabla_{(G)}^\nu(\r^2 \tensor{T}{_\nu_\r^{\dot\a}}) 
   = \frac 12 \r^2\sqrt{|G|}^{-1}
    \varepsilon^{\nu\s\r'\m}G_{\r\r'}  T_\mu 
   \tensor{T}{_\nu_{\s}^{\dot\a}}   
   +  m^2 \tensor{E}{^{\dot\a}_\r} \ ,
   \label{torsion-frame-eom-2}
\end{align}
or, using the on-shell relation \eq{T-del-onshell} for $T_\mu$,
 \begin{align}  
  \boxed{
 \nabla_{(G)}^\nu(\r^2 \tensor{T}{_\nu_\r^{\dot\a}}) 
   = \frac 12 \sqrt{|G|}^{-1}
    \varepsilon^{\nu\s\r'\m}G_{\r\r'}  \del_\mu\tilde\r \,
   \tensor{T}{_\nu_{\s}^{\dot\a}}   
   +  m^2 \tensor{E}{^{\dot\a}_\r} \ .
     }
   \label{torsion-frame-eom-3}
\end{align}
In terms of differential forms (see~\eqref{Bianchi-framelike}), this equation can be written concisely as
\begin{align}
\boxed{ \
  d(\r^2\star T^{\dot\a}) = d\tilde\r\wedge T^{\dot\a} + m^2 \star E^{\dot\a}  
  \ .\ }
  \label{eom-frame-form}
\end{align}
This can easily be verified for the cosmic background solution.
If the derivatives $\del\r \approx 0 \approx \del\tilde\r$ vanish or are negligible, this reduces to
the linear equation for the torsion
 \begin{align} 
  \nabla_{(G)}^\nu \tensor{T}{_\nu_\mu^{\dot\a}} 
  = \frac{m^2}{\r^2} \tensor{E}{^{\dot\a}_\mu} 
  \label{eom-torsion-noAS}
\end{align}
which has the structure of Maxwell equations for each frame index $\dot\a$.
Note that $\frac{m^2}{\bar\r^2} = \frac{3}{a^2(t)}$ is the cosmic curvature scale, so that 
the rhs is often negligible.
This equation can be rewritten as harmonic equation for the frame $\tensor{E}{^{\dot\a}_\nu}$
\begin{align}
    \Delta_{(G)}\tensor{E}{^{\dot\a}_\nu} 
   -\r^{-2}\cR_{\nu\s}\tensor{E}{^{\dot\a}^\s}
   = \frac{m^2}{\r^2} \tensor{E}{^{\dot\a}_\nu} \ ,
   \label{Laplace-E-explicit}
\end{align}
where $\Delta_{(G)} = \nabla_{(G)}^\mu \nabla^{(G)}_\mu $,
dropping again $\del\r \approx 0 \approx \del\tilde\r$ and
using the divergence constraint \eq{frame-div-free}. These are reminiscent 
of Maxwell equations for the vector potential.

\subsection{Einstein equation and geometric energy-momentum tensor}
\label{sec:Ricci-eq}

The Ricci tensor for the 
effective metric $G_{\mu\nu}$ can be expressed in terms of the 
torsion and the dilaton. 
Using  the e.o.m.~\eq{torsion-eq-nonlin-coord} for the torsion, the following relation was derived in \cite{Steinacker:2020xph}  
\begin{align}
\boxed{\
 \cR_{\mu\nu}  - \frac 12 G_{\mu\nu} \cR  = {\bf T}_{\mu\nu} 
  \ }
  \label{Einstein-eq-vac}
\end{align}
(absorbing a factor $8\pi$ for convenience)
where ${\bf T}_{\mu\nu}$ is the effective energy-momentum tensor associated to the torsion,
\begin{align}
  {\bf T}_{\mu\nu} &= - \frac 12 (\tensor{T}{_\r^{\d}_\nu}\tensor{T}{_\mu_\d^\r} 
  + \tensor{T}{_\r^{\d}_\mu}\tensor{T}{_\nu_\d^\r})
  - \tensor{K}{_\d^\r_\mu}\tensor{K}{_\r^\d_\nu}
  + 2\r^{-2}\del_\mu\r \del_\nu\r \nn\\
 &\quad + G_{\mu\nu} \Big(
 - \frac 14 \tensor{T}{^\s^\d^\r}\tensor{T}{_\d_\r_\s} 
 + \frac 18 \tensor{T}{^\d^\s^\r}\tensor{T}{_\d_\s_\r} 
  - \r^{-2} \del\r\cdot\del\r - 3R^{-2}\r^{-2} \Big) \ . 
  \label{em-tensor-effective}
\end{align}
As shown in  appendix \ref{sec:app-e-m-tensor} (see~\eq{e-m-torsion}), this can be written more succinctly as
\begin{align}
  {\bf T}_{\mu\nu}  &=  {\bf T}_{\mu\nu}[E^{\dot\a}] 
   + {\bf T}_{\mu\nu}[\r] 
    +  {\bf T}_{\mu\nu}[\tilde\r] 
       - \r^{-2} m^2 G_{\mu\nu} \ 
       \label{em-tensor-effective-explicit} 
\end{align}
in terms of contributions of  the dilaton $\r$, the axion $\tilde\r$,
and a Maxwell-like contribution from the frame fields $E^{\dot\a}$.
This shows again that the frame is physical, and acts -- together with the 
dilaton and the axion -- as source of the Einstein equations. 
The contributions 
${\bf T}_{\mu\nu}[E^{\dot\a}]$ and $ {\bf T}_{\mu\nu}[\tilde\r]$ 
of the frame and the axion turn out to have slightly non-standard form, and 
Ricci-flat geometries may arise if the various contributions cancel. 
While this may seem unlikely at first sight, it was shown previously that the 
standard Ricci-flat solutions arise indeed in the linearized regime \cite{Sperling:2019xar,Steinacker:2019dii}, and 
they are expected to arise more generally in the absence of axions and dilatons 
in view of \eq{Laplace-E-explicit}.

\section{An effective action}
\label{sec:eff-action}

Even though 
$E^{\da},\r$ and the metric $G_{\mu\nu}$ are mutually related, let us consider 
them as independent quantities for the moment.
Then all solutions of the above equations of motion 
are critical points of the following effective action
\begin{align}
  S_{\rm eff}[E,G,\r]
 &=  c_\cR S_\cR + c_E S_E + c_T S_{T} + c_\r S_\r 
 + c_m S_m + c_{\tilde m} S_{\tilde m} 
\end{align}
where 
\begin{align}
 S_\cR &=  \int\! d^4x \sqrt{|G|} \cR[G]   \nn\\
 S_E &= \int \! d^4x\sqrt{|G|}\r^{2}G^{\nu\nu'}G^{\s\s'} \tensor{T}{_\nu_{\s}^{\dot\a}}\tensor{T}{_{\nu'}_{\s'}_{\dot\a}} 
   = 2\int \r^2 dE^{\dot\a}\wedge \star d E_{\dot\a}   \nn\\
   S_{T} &= \int\! d^4x \sqrt{|G|} G^{\mu\nu}T_\mu T_\nu  \nn\\
 S_\r &=  \int\! d^4x \sqrt{|G|} G^{\mu\nu}\r^{-2}\del_\mu\r \del_\nu\r  \nn\\
 S_m &= \int\! d^4x\sqrt{|G|} m^2  \tensor{E}{^{\dot\a}_\k} \tensor{E}{_{\dot\a}_{\k'}} G^{\k\k'} \nn\\
S_{\tilde m} &= \int\! d^4x\sqrt{|G|} m^2 \r^{-2} \ .
 \label{S-eff}
\end{align}
Here $G, E$ and $\r$ are considered as independent objects which define the 
torsion $\tensor{T}{_\mu_{\nu}^{\dot\a}}$, and $T_\mu$ is defined 
as  $T=\star T^{(AS)}$ \eq{TAS-def}.
The variations of these terms are as follows
\begin{align}
 \d S_\cR &= \int\! d^4x \sqrt{|G|}  \Big(\cR_{\mu\nu} - \frac 12 G_{\mu\nu} \cR\Big) \d G^{\mu\nu} \nn\\[1ex]
 \d S_{E} 
   &= \int d^4x \sqrt{|G|} \big(-2 \d G^{\mu\nu} {\bf T}_{\mu\nu}[E^{\dot\a}] 
   -4\nabla^{\nu}(\r^{2} G^{\k\k'} \tensor{T}{_\nu_{\k}^{\dot\a}})\d\tensor{E}{_{\dot\a}_{\k'}}  \nn\\
  &\quad + 2 G^{\nu\nu'}\tensor{T}{_\nu_{\s}^{\dot\a}}\tensor{T}{_{\nu'}^{\s}_{\dot\a}} \r^{-1}\d\r \big) \nn\\[1ex]
  \d S_T 
   &=   \int \! d^4x\sqrt{|G|} \big(-2  \d G^{\mu\nu} {\bf T^{(AS)}}_{\mu\nu}[T] 
   + 4 T\cdot T \r^{-1}\d\r \big) \nn\\
 &\quad  - 2\int\! d^4x\, \r^2\d\tensor{E}{_{\dot\a}_\mu}
  \big( T_\k \varepsilon^{\nu\mu\s\k} \tensor{T}{_\nu_\s^{\dot\a}}
    + \tensor{E}{^{\dot\a}_\s}
   \varepsilon^{\nu\mu\s\k} \r^{-2}\del_\nu( \r^2  T_\k)\big) \nn\\[1ex]
\d S_\r
  &= \int\! d^4x \sqrt{|G|}\Big( \frac 12\d G^{\mu\nu} {\bf T_{\mu\nu}}[\r] 
   -2\nabla^{\mu}_{(G)} (\r^{-1}\del_\mu\r) \r^{-1} \d\r \Big) \nn\\[1ex]
 \d S_m &= \int\! d^4x \sqrt{|G|}\big(-  m^2 \r^{-2}G_{\mu\nu}\d G^{\mu\nu} 
+ 2 m^2 G^{\mu\nu} \tensor{E}{^{\dot\a}_\mu} \d\tensor{E}{_{\dot\a}_{\nu}}\big) \nn\\[1ex]
 \d S_{\tilde m} &= \int\! d^4x\sqrt{|G|}  \Big(-\frac 12 m^2\r^{-2} G_{\mu\nu}\d G^{\mu\nu}
 -2 m^2 \r^{-3} \d\r \Big)
 \label{variations-action-1}
\end{align}
using the results of Appendix \ref{sec:app-e-m-tensor}.
The coefficient of $\d \r$ coincides with the e.o.m.~\eq{Box-r-onshell} if 
\begin{align}
 c_\r = - 4c_E\ , \quad c_T =  c_E\ , \quad c_{\tilde m} = -8 c_E \ .
\end{align}
The coefficient of $\d G^{\mu\nu}$ coincides with the e.o.m.~\eq{Einstein-eq-vac} and using
\eq{em-tensor-effective-explicit} if 
\begin{align}
 c_E = \frac 12 c_\cR\ , \quad c_T = \frac 12 c_\cR\ , \quad  c_\r = -2 c_\cR\ ,
 \quad c_m =  c_\cR \ .
\end{align}
Finally, the coefficient of $\d E^{\dot\a}$ vanishes as a consequence of the e.o.m.~\eq{torsion-frame-eom-2} and \eq{eom-AS-T} if
\begin{align}
 c_T =  c_E \ .
\end{align}
Remarkably, all these conditions are compatible.
Therefore all critical points of the semi-classical matrix model  are critical points of 
\begin{align}
\boxed{\ 
 S_{\rm eff}[E,G,\r]
 =  2 S_\cR + S_E + S_{T} - 4 S_\r  + 2 S_m - 8 S_{\tilde m} 
  \ }
  \label{eff-action-EGr}
\end{align}
{\em for independent variations of $E,G$ and $\r$}. 
This result should  be useful to understand better the present gravity theory.
For example, it provides an ``explanation'' for (or at least a book-keeping of) the 
effective Einstein equation \eq{Einstein-eq-vac} and the 
explicit form of the contributions \eq{em-tensor-effective-explicit} on the rhs.

It  should be noted that the sign of  $S_E$ is non-standard, which is 
reflected in the negative sign in the energy-momentum tensor associated to 
the frame \eq{e-m-tensor-frame}. This looks ``wrong'' at first sight, but remember that the 
$E^{\dot\a}$ does not play the role of a vector field coupled to gravity,  it  rather
{\em defines} the metric. This is very different from the usual role of a vector field 
in gravity, and it was shown in \cite{Steinacker:2019awe} that no ghosts arise in this theory,
at least at the linearized level.

\paragraph{Extra gauge invariance.}

Note that the unrestricted action $S_{\rm eff}[E,G,\r]$ enjoys the gauge invariance
\begin{align}
\tensor{E}{^{\dot\a}_{\mu}} \to \tensor{E}{^{\dot\a}_{\mu}}  + \del_\mu\L^{\dot\a}
\label{extra-gaugeinv}
\end{align} 
subject to the constraint
\begin{align}
0 = \tensor{E}{^{\dot\a}^\mu} \del_\mu \L_{\dot\a}
\end{align}
(sum over $\dot\a$!),
which guarantees that the mass term $S_m$ is invariant. 
We shall not pursue this observation any further here.

\paragraph{Alternative action using the axion.}

Now we impose the e.o.m.~\eq{eom-AS-T} for the totally antisymmetric part of the torsion, 
replacing $T_\mu \to \r^{-2}\del_\mu\tilde\r$  according to \eq{T-del-onshell}.
If we  consider $\tilde\r$ as independent quantity, its e.o.m.~\eq{div-T-dEdE}
is recovered if we replace
the term $S_T$ by
\begin{align}
 S_T \to c_{\tilde\r} S_{\tilde\r} + c_{\tilde E} S_{\tilde E} 
\end{align}
where
\begin{align}
S_{\tilde\r} &=  \int \! d^4x \sqrt{|G|}G^{\mu\nu}\r^{-4}\del_\mu\tilde\r \del_\nu\tilde\r \nn\\
 S_{\tilde E} &= \int \! d^4x\tilde\r\,\varepsilon^{\nu\s\mu\k} \tensor{T}{_{\nu}_{\s}^{\dot\a}} \tensor{T}{_{\mu}_{\k}_{\dot\a}}
  \ = 4\int \tilde\r \, dE^{\dot\a}\wedge d E_{\dot\a}  \ .
\end{align}
Hence $\tilde \r$ is recognized as axion associated to the frame field.
The variations are 
\begin{align}
 \d S_{\tilde\r} &= \int\! d^4x\sqrt{|G|}\big(-2 \nabla^{\mu}_{(G)} (\r^{-4}\del_\mu\tilde\r)\d\tilde\r
   -4 G^{\mu\nu} \r^{-5}\del_\mu\tilde\r \del_\nu\tilde\r \d\r 
 + 2 {\bf T}^{\mu\nu}[\tilde\r] \d G^{\mu\nu}\big) \nn\\[1ex]
 \d S_{\tilde E} &= \int\! d^4x\big( 4\varepsilon^{\nu\s\mu\k} \tensor{T}{_{\nu}_{\s}^{\dot\a}} \del_\k\tilde\r \, \d\tensor{E}{_{\mu}_{\dot\a}} 
  - \varepsilon^{\nu\s\k\mu} \tensor{T}{_{\nu}_{\s}^{\dot\a}} \tensor{T}{_{\mu}_{\k}_{\dot\a}}\d\tilde\r \big) \ .
\end{align}
The $\d\tilde\r$ terms reproduce the e.o.m.~\eq{div-T-dEdE} for $\tilde\r$ 
if $c_{\tilde E} = -\frac 12 c_{\tilde\r}$.
The coefficient of $\d \r$ agrees with that of $S_T$ if $c_{\tilde\r} = - c_T$,
and the remaining equations of motion are also satisfied if 
$c_T = 2 c_{\tilde E}$. 
Therefore all critical points of the semi-classical matrix model are 
critical points of 
\begin{align}
\boxed{\ 
 S_{\rm eff}[E,G,\r,\tilde\r]
 =  2 S_\cR + S_E - S_{\tilde\r} + \frac 12 S_{\tilde E}  - 4 S_\r  + 2 S_m - 8 S_{\tilde m} 
  \ }
  \label{eff-action-EGrr}
\end{align}
{\em for independent variations of $E,G, \r$ and $\tilde\r$}. 
This is somewhat reminiscent of axion-dilaton gravity, cf.~\cite{Galtsov:1995zm,Bakas:1996dz,Clement:1996nh}, though 
the kinetic term for the frame is distinct.

\paragraph{Constrained action and triviality.}

Now we impose the constraint \eq{eff-metric}
\begin{align}
 G^{\mu\nu} = \r^{-2} \tensor{E}{^{\dot\a}^\mu}\tensor{E}{_{\dot\a}^\nu} \ .
 \label{G-frame-constraint}
\end{align}
Using the  identity \eq{R-offshell-T-2} which implies
 \begin{align}
  S_E   &= - 2S_{\cR} - S_T + 4 S_\r \ ,
        \label{R-offshell-action-id}
 \end{align}
the above action \eq{eff-action-EGr}
\begin{align}
 S_{\rm eff}[E,G,\r]
 =  2 S_\cR + S_E + S_{T} - 4 S_\r  + 2 S_m - 8 S_{\tilde m} 
\end{align}
turns out to be trivial,
\begin{align}
 S_{\rm eff}[E]
 &= 0
  \label{eff-action-EGrr}
\end{align}
noting that $S_m = 4 S_{\tilde m}$ when \eq{G-frame-constraint} is imposed.
Hence the effective action $S_{\rm eff}[E]$ is trivial
once the constraints are imposed. This may be somewhat disappointing, however it may
still provide a useful re-formulation of the system in terms of unconstrained quantities.

\section{Relation with the Cartan formalism and (no) local Lorentz transformations}
\label{sec:local-Lorentz-frame}

It is well-known that for any frame $\tensor{\cE}{_{\dot\a}_{\nu}}$ associated to a given metric $G_{\mu\nu}$,
one can act with a local Lorentz transformation on the frame as follows
\begin{align}
  \tensor{\cE}{_{\dot\a}_{\nu}}(x) 
  \to \tensor{\L}{_{\dot\a}^{\dot\b}}(x) \tensor{\cE}{_{\dot\b}_{\nu}}(x) \ 
\end{align}
leading to the same metric. In this sense
the frames form an $SO(3,1)$ bundle over space-time, leading 
to the Cartan formalism with a spin connection as discussed further below.
However, in the present setting, 
the frame $E_{\dot\b } = \r^{-1}\cE_{\dot\b }$ satisfies 
 the divergence constraint \eq{frame-div-free}, as well as the 
 on-shell relation \eq{eom-AS-T}. 
 This means that 
 the above local Lorentz (gauge) symmetry is broken, and the extra
 degrees of freedom of the frame encode the dilaton $\r$ and the axion $\tilde\r$.
 This is not a problem, but it exhibits a fundamental difference between 
  general relativity and the present theory, where the frame is a fundamental object.
  The dilaton $\r$ is also related to the invariant symplectic volume 
  via \eq{rho-M-G-relation}, leading to a reduction of the diffeomorphism invariance to 
  the volume-preserving diffeos \eq{diffeo-constraint}.
  
 We should therefore ask  the following question:
 Given some metric $G_{\mu\nu}$, can we always find a gauge 
 (i.e.\ a representative in the frame bundle)  and functions $\r$ and $\tilde\r$
 such that  $E_{\dot\b } = \r^{-1}\cE_{\dot\b }$ satisfies 
 the divergence constraint \eq{frame-div-free} as well as the 
 conditions \eq{T-del-onshell}?
By counting degrees of freedom (d.o.f.), it is plausible that the answer 
should be generically yes.
Indeed, if we consider $\r$ and $\tilde\r$ as independent fields, then 
the 6 d.o.f.\ of $\tensor{\L}{_{\dot\a}^{\dot\b}}(x)$ together with the 
2 d.o.f.\ $\r$ and $\tilde\r$ should allow 
to satisfy the 4+4 equations \eq{T-del-onshell} and  \eq{frame-div-free}. 
But this means that both the dilaton $\r$
and the axion $\tilde\r$ are determined by the  given 
metric $G_{\mu\nu}$. This is reflected in relations such as \eq{R-axion-r-onshell}.
In particular they do not add any propagating degrees of freedom.
This is consistent with the results in \cite{Steinacker:2019awe,Sperling:2019xar}
for the linearized case, where it was shown that the only physical, propagating modes
are those encoded in the degrees of freedom of a (massive) graviton.
A more detailed understanding of $\r$, $\tilde\r$ and their relation 
to the geometry should be developed elsewhere.

\paragraph{Relation with the Cartan structure equations.}

To make contact with the standard Cartan formalism of general relativity, 
we consider the  (co-)frame \eq{eff-frame} 
$\cE_{\dot\b } = \r E_{\dot\b } = \tensor{\cE}{_{\dot\b}_\mu} dx^\mu$. Then
the  spin connection 
one-form $\tensor{\omega}{_{\dot\a}_{\dot\b}} = \tensor{\omega}{_\mu_{\dot\a}_{\dot\b}} dx^\mu
 = - \tensor{\omega}{_{\dot\b}_{\dot\a}}$ satisfies
 the first Cartan structure equations
\begin{align}
  d \cE_{\dot\a} &= -\tensor{\omega}{_{\dot\a}^{\dot\b}}\wedge \cE_{\dot\b}  \ .
 \end{align} 
 Clearly this is closely related to the 
 (con)torsion of the Weitzenb\"ock connection. Indeed, 
 \begin{align}
 d \cE_{\dot\a} =\frac 12 (\del_\mu \tensor{\cE}{_{\dot\a}_{\nu}} - \del_\nu \tensor{\cE}{_{\dot\a}_{\mu}})dx^\mu\wedge dx^\nu
   &= -\tensor{\omega}{_\mu_{\dot\a}^{\dot\b}} \tensor{\cE}{_{\dot\b}_{\nu}} dx^\mu\wedge dx^\nu
   = \frac 12 \cT_{\mu\nu{\dot\a}}dx^\mu\wedge dx^\nu
 \end{align}
so that 
\begin{align}
 \cT_{\mu\nu{\dot\a}} &= 
 -\tensor{\omega}{_\mu_{\dot\a}^{\dot\b}} \tensor{\cE}{_{\dot\b}_{\nu}}
 +\tensor{\omega}{_\nu_{\dot\a}^{\dot\b}} \tensor{\cE}{_{\dot\b}_{\mu}}
  =  \tensor{\omega}{_\nu_{\dot\a}_\mu} -\tensor{\omega}{_\mu_{\dot\a}_\nu} \ .
  \label{T-omega-relation}
\end{align}
This provides the relation of the spin connection $\omega$ of the frame bundle
to the  torsion tensor $\cT$ of the Weitzenb\"ock connection.
Even though the spin connection is not a tensor in the Cartan 
formalism due to the local Lorentz transformations, $\cT_{\mu\nu{\dot\a}}$ 
{\em is} a tensor 
in the present formalism, where the frame is physical.
Hence  $\cT_{\mu\nu{\dot\a}}$ can be viewed as a physical 
manifestation of  the 
spin connection in the Cartan formulation of Riemannian geometry.

\section{Spherically symmetric static solution}
\label{sec:solution}

In this section, we will find a simple rotationally invariant static 
solution of the  nonlinear equations of 
motion for the frame.

\paragraph{$SO(3)$-invariant frames.}
We are interested in local perturbations of the background which are
centered at $x^i = 0$, hence at $r=0$, 
and invariant under (global) $SO(3)$ rotations.
We first observe that
the background frame $\tensor{\bar E}{_{\dot\a}^\mu} = \sinh(\eta)\d^\mu_{\dot\a}$
\eq{frame-BG}
is  $SO(3)$-invariant if the frame index $\dot\a$ is transformed as a vector.
This is the manifest global $SO(3)$ symmetry of the matrix model, and
it seems natural to keep this  $SO(3)$  symmetry manifest for the perturbed 
rotationally-invariant geometry. 
This is easily achieved in Cartesian coordinates $x^\mu$, 
adopting the  notation\footnote{The $t$ defined here should not be confused with the comoving time that was introduced in section~\ref{sec:backgroundgeometry} and which will not be used in the current section.} 
\begin{align}
 t = x^0 \ , \qquad r^2 = x^i x^j \d_{ij} \ .
\end{align}
Then the most general spherically symmetric ansatz for the frame 
$\tensor{E}{^{\dot\a}_\mu}$ is 
\begin{align}
\tensor{E}{^0_0} &= A\ ,\nonumber\\
\tensor{E}{^i_0} &= Ex^{i}\ ,\nonumber\\
\tensor{E}{^0_i} &= Dx^{i}\ ,\nonumber\\
\tensor{E}{^i_j} &= Fx^{i}x^{j} + \delta^{i}_{\ j}B + S\epsilon_{ijm} x^{m} \ ,
\label{ansatz}
\end{align}
where $A,B,D,E,F$ and $S$ are assumed to be functions of $r$ only.
In particular we consider only static\footnote{More precisely, the configurations 
are static on scales much shorter than the cosmic expansion rate.} configurations.
We can eliminate $D$ and $F$ using a simple change of coordinates 
$t\to f(r) + t$
and $x^i \to g(r) x^i$, which is 
understood from now on.
In terms of differential forms $\tensor{E}{^{\dot\a}}= \tensor{E}{^{\dot\a}_\mu} dx^\mu$,
the (co)frame is then
\begin{align}
\tensor{E}{^0} &= A dt\ ,\qquad 
\tensor{E}{^i} = B dx^i + E x^i dt + S\epsilon_{ijm} x^{m} dx^j
\end{align}
and the associated torsion 2-form $T^{\dot\a} = dE^{\dot\a}$ is obtained as
\begin{align}
T^0 &= d E^0 =  A' dr \wedge dt\ , \nn\\
T^i &= d E^i = B' dr  \wedge dx^i + E d x^i  \wedge dt + x^i E' dr  \wedge dt 
+ S\epsilon_{ijm} d x^{m}  \wedge dx^j + S' \epsilon_{ijm} x^{m} dr  \wedge dx^j \ .
\label{ansatz-E-simple}
\end{align}
It is easy to see that the totally antisymmetric part 
$T^{(AS)}$ of the torsion vanishes if $S=0$, 
which we assume henceforth.
Then 
\begin{align}
 T_\mu = 0 \ , \qquad \tilde\r = const\ ,
\end{align}
and the frame is given 
in matrix form by 
\begin{align}
 \tensor{E}{^{\dot\a}_\mu} 
 =  \begin{pmatrix}
    A & 0 & 0 & 0 \\
    E x^1 & B & 0 & 0\\
    E x^2 & 0 & B & 0\\
    E x^3 & 0 & 0 &  B
   \end{pmatrix} \ .
\end{align}
The inverse frame is 
\begin{align}
 \tensor{E}{_{\dot\a}^\mu} 
 =  \begin{pmatrix}
    A^{-1} & -\frac{E}{AB} x^1  & -\frac{E}{AB} x^2 & -\frac{E}{AB} x^3 \\
    0 & B^{-1} & 0 & 0\\
    0 & 0 & B^{-1} & 0\\
    0 & 0 & 0 &  B^{-1}
   \end{pmatrix}
\end{align}
and the metric obtained from this frame reads
\begin{align}
&\g_{00} = -A^2 + r^2 E^2\ ,\nonumber\\
&\g_{0i} = B E x^{i}\ ,\nonumber\\
&\g_{ij} = \delta_{ij}B^2 \label{components1}
\end{align}
in Cartesian coordinates.
Now we can compute the dilaton $\rho$.
The condition \eq{tilde-T-T-contract} yields
\begin{align}
 -\frac{2}{\rho}\partial_{\mu}\rho =
\tensor{T}{_\mu_\nu_\da}\tensor{E}{^\da^\nu} 
= \partial_{\mu}\tensor{E}{^0_0}\tensor{E}{_0^0} 
+ \partial_{\mu}\tensor{E}{^i_j}\tensor{E}{_i^j} 
- \partial_{i}\tensor{E}{^0_\mu}\tensor{E}{_0^i} - \partial_{j}\tensor{E}{^i_\mu}\tensor{E}{_i^j} \ ,
\label{T-contract-2}
\end{align}
taking into account the above form for the (co)frame.
For the space-like components $\mu=k$, this gives 
\begin{align}
 -\frac{2}{\rho}\partial_{k}\rho 
 = A^{-1} \partial_{k} A + 2B^{-1}\partial_{k} B  
\label{T-contract-k}
\end{align}
so that $\r$ is determined by
\begin{align}
AB^2\rho^2= c_{1}
\label{rho-explicit}
\end{align}
where $c_{1}$ is a numerical constant. For the time-like components $\mu=0$, 
\eq{T-contract-2} leads to 
\begin{align}
 -\frac{2}{\rho}\partial_{0}\rho 
&= 0 = \frac{1}{B}\big(r E A^{-1} A' -  (r E' + 3E)  \big)
\end{align}
which is solved by
\begin{align}
  c_2 A r^{-3} &= E \ .
 \label{E-constraint}
\end{align}
In particular, we can set $E=0$ for $c_2=0$.
Furthermore, one finds for the determinants
\begin{align}
\det (\g_{\mu\nu})  = -A^2 B^6 , \qquad 
\sqrt{|G|} 
= \frac{c_1^2}{AB}
 \label{rho-detG}
\end{align}
where $E$ drops out. In particular, 
we observe 
\begin{align}
 \sqrt{|G|} \r^{-2} 
  = c_1 B \ .
\end{align}
It is straightforward to check that these relations also imply the 
divergence constraint \eq{frame-div-free}.
The components of the effective metric $G_{\mu\nu} = \r^2\g_{\mu\nu}$  are
then obtained explicitly as
\begin{align}
&G_{00} = c_1\Big(-\frac A{B^2}+ r^2 \frac{E^2}{AB^2}\Big)
 = c_1(-1+ c_2^2 r^{-4} )\frac A{B^2}\ ,
\nonumber\\
&G_{0i} =  c_1 \frac E{AB} x^{i} 
 = c_1 c_2 \frac {1}{AB} r^{-3} x^{i}  \ ,\nonumber\\
&G_{ij} =  c_1 \frac 1{A}\delta_{ij} \ .
\label{components-G-1}
\end{align}
We will focus on the case $E=0$ (and $S=0$) in this paper for simplicity, 
so that $c_2=0$.
Then  
\begin{align}
ds_G^{2} 
= -c_1\frac A{B^2} d t^2
 + c_1 \frac 1{A}\delta_{ij} dx^i dx^j \ .
\label{metric-cartes-2}
\end{align}

\paragraph{Equations of motion and solution.}

For the spherically invariant frames as above, it is  convenient to use the 
formulation \eq{eom-frame-form} of the equations of motion. Using
\begin{align}
 \star T^{\dot 0} &= r^{-1} A' x^i \star(dx^i\wedge dt) 
  = r^{-1} A' \sqrt{|G|} G^{ii'}G^{00} \varepsilon_{i'kl}x^i  dx^k\wedge dx^l  \nn\\
 \star T^{\dot k} &= r^{-1} B' x^i \star(dx^i \wedge dx^k)  
  = r^{-1} B' \sqrt{|G|} G^{ii'}G^{kk'} \varepsilon_{i'k'l}x^i dt \wedge dx^l \ 
\end{align}
in Cartesian coordinates (sum over $i$ is understood),
we obtain
\begin{align}
 d(\star\r^2 T^{\dot 0}) &= r^{-1} A' \r^2\sqrt{|G|} G^{ii'}G^{00} \varepsilon_{i'kl}
  d x^i \wedge  dx^k\wedge  dx^l  \nn\\
 &\quad + (r^{-1} A' \r^2\sqrt{|G|} G^{ii'}G^{00} )' \varepsilon_{i'kl}x^i dr\wedge  dx^k \wedge dx^l  \nn\\
  d(\star \r^2 T^{\dot k}) 
   &= r^{-1} B' \r^2\sqrt{|G|} G^{ii'}G^{kk'} \varepsilon_{i'k'l}dx^i \wedge  dx^l \wedge dt  \nn\\
 &\quad  + r^{-1}(r^{-1}B'\r^2 \sqrt{|G|} G^{ii'}G^{kk'})'
   \varepsilon_{i'k'l}x^i x^j dx^j\wedge  dx^l \wedge  dt
\end{align}
where ${}'$ denotes radial derivative, and $G^{ij} \equiv A(r)\d^{ij}$.
Assuming $m=0$ for simplicity (as well as $S =0 = E$)  and using 
$\varepsilon_{ikl}x^i r dr \wedge dx^k\wedge  dx^l = 2 r^2 d^3x$ and $\varepsilon_{ikl}
  d x^i \wedge  dx^k\wedge  dx^l = 6 d^3x$,
the equations of motion  for $\dot\a=0$ reduce to 
\begin{align}
0 
 &= 2r^{-1} (A' \r^2\sqrt{|G|} A G^{00}) 
 + (A' \r^2\sqrt{|G|} A G^{00} )'   \nn\\
 0 &= \frac{d}{dr}\big(r^2\sqrt{|G|}\r^2 A G^{00}  A'\big) \ .
\end{align}
This gives 
\begin{align}
 \frac{d}{dr}\big(r^2 B^{-1} (A^{-1})'\big) = 0 \ . 
 \label{eom-A-firstorder}
\end{align}
On the other hand, the space-like equations  for 
$\dot\a = 1,2,3$ have no solutions unless $B'=0$; the same conclusion is reached using 
the formulation \eq{torsion-frame-eom-3}.
Thus assume $B(r) =b_0 = const$. Then we obtain 
\begin{align}
  (A^{-1})' = -\frac{M}{r^2}, \qquad 
 A^{-1} = 1+\frac Mr
\end{align}
for some constant $M$.
Here we imposed the asymptotic behavior\footnote{Strictly speaking we should introduce another normalization constant to recover the  
cosmic background frame for $r\to\infty$, which we drop for simplicity.}
\begin{align}
 A(r) \to 1 \qquad \mbox{for} \quad r\to\infty
\end{align}
corresponding to an asymptotically constant frame
\begin{align}
 \tensor{E}{^{\dot 0}} &= A(r) dt = \frac 1{1+\frac Mr} dt \ ,\qquad 
 \tensor{E}{^{\dot k}} = b_0 dx^k 
 \label{frame-rot-solution}
\end{align}
i.e.\ $\tensor{E}{^{\dot 0}_{0}} = A(r), \ \tensor{E}{^{\dot k}_{k}} =  b_0$.
This leads to the following non-trivial metric and dilaton
\begin{align} 
ds^2_G = G_{\mu\nu} dx^\mu dx^\nu &= - \frac{c_1 b_0^{-2}}{(1+\frac Mr)} dt^2 + c_1\Big(1+\frac Mr\Big)\sum_i (dx^i)^2 \nn\\
 &= - \frac{c_1 b_0^{-2}}{(1+\frac Mr)} dt^2 + c_1\Big(1+\frac Mr\Big)(dr^2 + r^2 d\Omega^2)   \nn\\
 \r^2 
  &= c_1 b_0^{-2} \Big( 1+\frac Mr \Big)\ .
\end{align}
This reproduces the linearized Schwarzschild metric 
(cf.~\cite{Steinacker:2019dii}),
but it deviates from the full Schwarzschild metric  at the 
non-linear level; this should be expected due to the dilaton. 
The same result can also be obtained from \eq{torsion-frame-eom-3}
using Gauss' theorem which yields \eq{eom-A-firstorder}, 
or directly from \eq{torsion-frame-eom-3} using the Christoffel symbols.
The apparent singularity of the metric at the origin is a coordinate artifact, and 
the metric is seen  (e.g.\ using $u=\sqrt{r}$) to be regular at the origin. 
We can compute the associated Ricci tensor $\cR_{\mu \nu}$ and the Einstein tensor $\cG_{\mu \nu}$: using the relations 
\begin{align}
 \r^2 G_{00} = -\a^2 \ , \qquad \a := c_1 b_0^{-2}
 \label{rho-G00-relation}
\end{align}
we find the following result in spherical coordinates 
for $c_1=1$  
\begin{align}
 \cR_{\mu\nu} dx^\mu dx^\mu &= \frac{M^2}{2(M+r)^2} d\Omega^2 
 + \frac{\a M^2}{2(M+r)^4} dt^2  \nn\\
 \cG_{\mu\nu} dx^\mu dx^\mu &=  -\frac{M^2}{4r^2 (M + r)^2} dr^2 
 +  \frac{M^2}{4(M + r)^2} d\Omega^2 + \frac{3 \a M^2}{4(M+r)^4} dt^2 \ .
 \label{Ricci-einstein-solution}
\end{align}
The Einstein tensor is non-vanishing,  but decays like $M^2 r^{-4}$ as $r\to\infty$, 
reflecting the linearized Schwarzschild geometry.\footnote{
Note that we must set $\a=1$
if the frame should reduce to the background frame for $r\to\infty$.}
This is clearly integrable at the origin, so that the solution
should be viewed as a vacuum or ``remnant'' solution, rather than representing the 
gravitational field of  matter at the origin. This is possible because the present 
gravity theory is richer than GR, and the present solution has a non-trivial dilaton. 
Indeed \eq{R-axion-r-onshell} implies that Ricci-flat solutions can arise  
(in the limit $m^2\to 0$) only if both $\r$ and $\tilde\r$ vanish or cancel each other. 
It is therefore not surprising that the present solution
differs from the Schwarzschild solution. 
However, we expect that there are other $SO(3)$-invariant  solutions with different characteristics. In particular, such solutions might contribute to the 
physics underlying the apparent ``dark matter''.

Finally,
it is interesting to note that the total ``apparent mass'' 
is proportional to $M$,
\begin{align}
 \int_0^\infty dr 4\pi r^2 \cG_{00} \   = \pi M \ .
\end{align}

\paragraph{Effective energy-momentum tensor \eq{em-tensor-effective} and Einstein equations.}

The above frame has a  simple diagonal structure, similar to the background frame.
It is related to the dilaton via
\begin{align}
 \tensor{E}{^{\dot 0}_{0}} = A(r) = \a\, \r^{-2} \ 
\end{align}
using \eq{rho-explicit} and \eq{rho-G00-relation}.
This leads to the following expressions for the frame-valued torsion
\begin{align}
 T^{\dot k} &= 0 \nn\\
 T^{\dot 0} &= A'(r) dr\wedge dt = -\a\,\r^{-4} d\r^2\wedge dt
 = \frac 1{(1+\frac Mr)^2} \frac{M}{r^2 } dr \wedge dt \ ,
\end{align}
so that the non-vanishing components are
\begin{align}
 \tensor{T}{_i_0^{\dot 0}} = - \tensor{T}{_0_i^{\dot 0}} 
 = -2\a\,\r^{-3} \del_i\r \ .
\end{align}
We note that only the time component $E^{\dot 0}$
of the frame contributes to the torsion,
with asymptotic behavior
\begin{align}
 T^{\dot 0} &\sim \frac 1{r^2} \to 0, \qquad r\to\infty \nn\\
  T^{\dot 0} &\to \frac 1M \neq 0, \qquad r\to 0 \ .
\end{align}
The energy-momentum tensor of the dilaton  \eq{e-m-tensor-rho}  is
\begin{align}
  {\bf T}_{\mu\nu}[\r] &= 2\r^{-2}\Big(\del_\nu\r \del_\mu\r - \frac 12 G_{\nu\mu}\del^\s\r \del_\s\r \Big)\ , \nn\\
   {\bf T}[\r] &= -2\r^{-2} \del^\s\r \del_\s\r = -2 \r^{-2} G^{rr}\del_r\r \del_r\r \ , \nn\\
  {\bf T}_{00}[\r] &= - G_{00}\r^{-2} \del^\s\r \del_\s\r 
   = \frac 12 G_{00} {\bf T}[\r] \ .
   \label{eom-rho-spheric}
\end{align} 
Hence the contribution of the frame to the energy-momentum tensor \eq{e-m-tensor-frame} is 
\begin{align}
 {\bf T}_{\mu\nu}[E^{\dot 0}] 
 &=  \r^2\big(\tensor{T}{_\mu_\s^{\dot0}}\tensor{T}{_\nu_\r^{\dot0}}G^{\r\s} 
  - \frac 14 G_{\nu\mu}(\tensor{T}{_\s_\k^{\dot0}}
  \tensor{T}{_{\s'}_{\k'}^{\dot0}}G^{\s\s'} G^{\k\k'})  \big) \nn\\
  &= 4\a^2 \r^{-4}
   \Big(G^{00}\big(\del_\mu\r \del_\nu\r 
    - \frac 12 G_{\nu\mu}\big(\del_\s\r \del_{\s'}\r G^{\s\s'}\big)  \big)
    + \d_\mu^0\d_\nu^0(\del_\s\r\del_{\s'} \r G^{\s\s'}) \Big)\nn\\
   &= 2\a^2 \r^{-2} 
    \big(G^{00}  {\bf T}_{\mu\nu}[\r] - \d_\mu^0\d_\nu^0{\bf T}[\r] \big) \nn\\
   &= - 2 {\bf T}_{\mu\nu}[\r] + 4\d_\mu^0\d_\nu^0 {\bf T}_{00}[\r] 
\end{align}
using \eq{rho-G00-relation}, \eq{eom-rho-spheric} and $\r=\r(r)$.
Note that the frame and the dilaton contribute the same terms with opposite sign, and
the total contribution to the Einstein equations is
given by
\begin{align}
 {\bf T}_{\mu\nu} 
 &= {\bf T}_{\mu\nu}[E^{\dot 0}] + {\bf T}_{\mu\nu}[\r] 
  = - {\bf T}_{\mu\nu}[\r] + 4\d_\mu^0\d_\nu^0 {\bf T}_{00}[\r]  \ .
\end{align}
Explicitly, for the above solution with  $c_1=1$ we obtain 
\begin{align}
  {\bf T} [\r] &= -\frac 12  \r^{-4}G^{rr}\del_r\r^2\del_r\r^2
  = - \frac 12 \frac{1}{(1+\frac{M}{r})^3}\frac{M^2}{r^4} 
\end{align}
hence
 \begin{align} 
  {\bf T}_{00} &=  3 {\bf T}_{00} [\r] = \frac 32 G_{00} {\bf T} [\r] 
    = \frac 34 \frac{\a M^2}{(r + M)^4} \nn\\
  {\bf T}_{rr}  &= - {\bf T}_{rr}[\r]
  = \frac 12 G_{rr} {\bf T} [\r] = -\frac 14 \frac{M^2}{(r+M)^2 r^2}  \nn\\
  {\bf T}_{\vartheta\vartheta}
  &= - {\bf T}_{\vartheta\vartheta}[\r]
  = -\frac 12 G_{\vartheta\vartheta} {\bf T} [\r] 
  = \frac 14 \frac{M^2}{(r+M)^2} \ .
  \label{e-m-tensor-solution}
\end{align}
This agrees precisely with the above results for the Einstein tensor \eq{Ricci-einstein-solution},
as it should.
Note that ${\bf T}_{\mu\nu} \sim M^2 r^{-4}$ as $r\to\infty$.
We will give an explicit solution $Z^{\dot\a}$ below, which realizes this
frame via the Poisson brackets.

\section{Semi-classical reconstruction of the matrices $Z_{\dot\a}$}
\label{sec:reconstruct}

Given a classical solution $\tensor{E}{_{\dot\a}^\mu}$ of the 
above equations for the frame, we must finally find a
corresponding solution $Z_{\dot\a}$ of the matrix model
such that
\begin{align}
  \{Z_{\dot\a},x^\mu\} = \tensor{E}{_{\dot\a}^\mu} \ .
  \label{frame-Poisson}
\end{align}
This problem was partially 
solved in \cite{Steinacker:2019awe} sections 6.3 (point 1) and 9.2,
where it was shown that the divergence constraint
\begin{align}
 \del_\mu(\r_M E^\mu) = 0
\end{align}
is satisfied for all $\{x_\mu,Z^{(1)}\}_0 \equiv \cA^{(-)}_\mu[Z^{(1)}]$ modes 
with $Z^{(1)} \in \cC^1$ (in the notation of \cite{Steinacker:2019awe}, cf.~\eq{cC-spin-decomp}),
and moreover these modes 
are complete in $\cC^0\otimes\R^4$ (together with the pure gauge modes). 
Here $(\,)_0$ denotes the 
projection to the spin 0 sector $\cC^0$, i.e.\ to functions on $\cM^{3,1}$. 
This means there is always some $Z_{\dot\a}\in\cC^1$ such that 
$\{Z_{\dot\a},x^\mu\}_0 = \tensor{E}{_{\dot\a}^\mu}$ as desired, but
this may be accompanied by higher spin contributions 
$\{Z_{\dot\a},x^\mu\}_2 \in\cC^2$. Whether or not these higher-spin 
contributions can be canceled by suitable higher-spin ``counterterms'' 
or coordinate re-definitions $\tilde x^\mu$ is  an open question in general.

We shall illustrate this reconstruction explicitly for the  spherically 
invariant solution \eq{frame-rot-solution}, by providing explicit functions $Z^{\dot \a}$ on the
bundle space. In that case  it is indeed possible to cancel the higher-spin contributions (for the dominant terms at late times), 
but this requires an infinite tower of 
higher-spin modes from the point of view of the cosmic background.

\paragraph{Semi-classical $Z^{\dot\a}$  for the spherical solution.}

For the specific solution \eq{frame-rot-solution},
we should accordingly find ``potentials'' $Z^{\dot\a}$ 
-- and possibly new coordinates $\tilde x^\mu$ -- such that
\begin{align}
\begin{array}{lll}
 &\{Z^0, \tilde x^0\} = -\sinh(\eta) f(r^2)\ , \qquad
 & \{Z^0, \tilde x^i\} \approx 0   \nn\\
 &\{Z^i,\tilde x^0\} \approx 0 \ ,  &\{Z^i,\tilde x^j\} \approx  \sinh(\eta) \d^{ij}
 \end{array}
\end{align}
for $f(r) = A^{-1}(r)$.
Here we re-inserted a factor $\sinh(\eta)$ to recover the background frame 
\eq{frame-BG}.
The second line suggests to leave the space-like generators $Z^k = t^k$ and the $x^i$ coordinates unchanged. Then
$x^0$ should  not be modified either, in order to preserve $\{Z^i,x^0\}=0$. 
The first ansatz one might try is $Z^0 = t^0 g(r)$ for some  function $g(r^2)$.
This would give
\begin{align}
\{g(r^2)t^0,x^0\} 
  &= g(r^2)\{t^0,x^0\} + t^0\{g(r^2),x^0\} \nn\\
  &= -\sinh(\eta)\Big(g(r^2) + 2 g'(r^2) p_0^2\Big) \ \  \in \cC^0 \oplus \cC^2 
  \label{simple-ansatz-g}
\end{align}
using \eq{invar-brackets}.
However,
the component in $\cC^2$ is not subleading and hence cannot be neglected
since  $|p_0|^2 = |\frac{p\cdot x}{x^0}|^2 = r^2$, cf.\ section 
\ref{sec:so3-inv-structure}.
Therefore we consider the more general ansatz  
\begin{align}
\begin{array}{lll}
 & Z^0 :=  t^0 g(r,\chi)\ , \qquad  & Z^k = t^k  \nn\\
 & \tilde x^0 = x^0\ , \qquad & \tilde x^i = x^i
  \end{array}
 \label{perturbed-ansatz}
\end{align}
where $\chi$ is the central generator in the $SO(3)$-invariant sub-algebra  given in \eq{chi-def},
which allows for higher spin corrections.
Then 
\begin{align}
 \{t^0 g(r^2,\chi),x^0\} 
  &=  -\sinh(\eta) g(r^2,\chi) - 2 \tilde R x^4\del_{r^2} g(r^2,\chi) t^0p^0 \nn\\
  &=  -\sinh(\eta) g(r^2,\chi) - \sinh(\eta) \del_{r^2} g(r^2,\chi) 2(p^0)^2 \nn\\ 
  &=   -\sinh(\eta) \big(1 + 2(r^2-\chi)\del_{r^2} \big)g(r^2,\chi)
\end{align}
so that we need to solve 
\begin{align}
 \big(1 + 2(r^2-\chi)\del_{r^2}\big) g(r^2,\chi) = f(r^2) \ .
\end{align}
The solution is\footnote{Here $p^0 = \pm\sqrt{r^2 - \chi}$ \eq{p0-r-chi} is understood. The explicit $\frac{1}{p_0}$
cancels in $\tilde Z^0$, so that the sign ambiguity drops out. Note that $p^0$ has a clean definition 
as $\msu(4,2)$ generator at the matrix level, hence this is not a problem.}
\begin{align}
 g(r^2,\chi) 
  = \frac{1}{p_0}\Big(a(\chi) + \frac 12\int_\chi^{r^2} \frac{f(u)}{\sqrt{u-\chi}} du\Big) \ .
  \label{g-solution}
\end{align}
The term with $a(\chi)$  satisfies the homogeneous equation, hence we can simply drop $a(\chi)$,
or use it to impose boundary conditions.
For the above solution with 
\begin{align}
 f(r^2) = A^{-1} = 1+\frac{M}{r}
\end{align}
(assuming $\sinh(\eta) \approx {\rm const}$),
the above integral can be evaluated explicitly,
\begin{align}
 \int_\chi^{r^2} \frac{1+\frac{M}{\sqrt{u}}}{\sqrt{u-\chi}} du
 &= 2 \left(M \log \left(r +\sqrt{r^2-\chi}\right)+\sqrt{r^2-\chi}\right)-M \log (\chi)  \nn\\
  &= 2 \left(M \log \left(r + p_0\right)+p_0\right)- M \log (r^2-p_0^2) \nn\\
  &= 2p_0 +  M\log \Big(\frac{r + p_0}{r-p_0}\Big) \ .
\end{align}
We recall that  $\frac{p_0}{r} \approx \frac{\vec p\cdot \vec x}{|\vec p||\vec x|} =: \cos\vartheta$ and therefore  $-r \leq p_0 \leq r$  due to  \eq{scalar-quant-relations}  and \eq{p0-bounds}, so that the last term 
is well-defined up to  a mild log-type singularity  
 at antipodal points  $\vartheta = 0,\pi$ of the internal $S^2$ fiber.
Thus
\begin{align}
\boxed{\ 
 \tilde Z^0 = t^0 + \frac{M}{2R\tilde R} \log \Big(\frac{r + p_0}{r-p_0}\Big) 
  = t^0 + \frac{M}{2R\tilde R} \Big(\frac{2p_0}{r} + \frac 23 \frac{p_0^3}{r^3} + ...\Big) 
   = t^0 A^{-1}(r) + O\Big(\frac{p_0^3}{r^3}\Big) \ ,
  \ }
\end{align}
which has indeed the structure in \eq{simple-ansatz-g}  up to higher-spin corrections. 
One can check directly that it satisfies the desired relation 
\begin{align}
\boxed{\ 
 \{\tilde Z^0, x^0\} = -\sinh(\eta) \big(1+\frac{M}{r}\big) = -\sinh(\eta)A^{-1}(r) \ .
\ }
\end{align}
Finally, note that $\{\tilde Z^0,x^i\} = p^0 \{g,x^i\} \ll \sinh(\eta)$ is negligible at late times
compared to $\{\tilde Z^0,x^0\}$,
because  $p^0, r^2,\chi$ and $\theta^{ij}$  are all bounded i.e.\ they do not grow with $\eta$. Therefore the $Z^{\dot\a}$ indeed reconstruct the frame at late times 
 $\eta\gg 1$.

The above solution certainly makes sense as a formal power series,\footnote{cf.\ the black hole solutions in Vasiliev higher 
spin theory \cite{Didenko:2009td,Iazeolla:2011cb}.}
but  we should require that $\tilde Z^0$ exists as a function on the bundle space $\C P^{1,2}$.
As such, the above solution $\tilde Z^0$ exhibits a logarithmic singularity for $p_0 = \pm r$,
which arises at the antipodes of the internal $S^2$ where $\vec p \parallel \vec x$. One of them can be canceled by choosing a suitable $a(\chi)$
in \eq{g-solution}, but not simultaneously for both singularities. 
However, this singularity is integrable. This strongly suggests that the solution 
is acceptable, since the underlying framework of quantized symplectic spaces 
is based on the relation $End(\cH_n) \sim L^2(\C P^{1,2})$ between Hilbert-Schmid operators
and square-integrable
functions on $\C P^{1,2}$. 
At the operator level,  the internal $S^2$ is actually a fuzzy sphere \cite{Sperling:2018xrm} and $p^0$ arises from
a $\msu(4,2)$ generator $P^0$, where the semi-classical bound \eq{p0-bounds} is 
replaced by a strict operator bound.
All this strongly suggests that there should be a well-defined underlying 
matrix solution, which  should  be studied in more detail elsewhere.

The situation at $r=0$ is more intricate. Then the location of the singularity on $S^2$ 
becomes inconsistent, which signals a more serious singularity.
This suggests that 
some extra structure is present at the origin. On the other hand, 
the energy-momentum tensor  \eq{e-m-tensor-solution} is integrable at the origin.
This suggests that the solution should be viewed as
a purely geometric matrix configuration with a ``quantum'' or matrix singularity
at the origin. Such a matrix solution might be considered as a ``remnant'', which
is precisely the type of structure where the underlying matrix framework may provide 
unique new insights.

\section{Discussion}

We have elaborated the non-linear dynamics of the effective gravity sector which emerges 
from the IKKT or IIB matrix model on a certain type of covariant quantum space.
This should be viewed as a candidate for a modified gravity theory.
The most convenient description seems to be in terms of the frame $E^{\dot\a}$,
for which we find a simple covariant equation of motion. This captures 
the equation of motion derived previously for the torsion, avoiding the use of the
Weitzenb\"ock connection.

However, there are important differences to 
the frame formulation of general relativity. The main difference is that 
 the local Lorentz invariance of the frame bundle is broken, and
the frame contains more information than just the  metric. All frames which 
arise in the model have the form $E^{\dot\a} = \{Z^{\dot\a},.\}$,
which entails a divergence constraint. This is consistent with the fact that 
diffeomorphism invariance is reduced to a type of volume-preserving diffeomorphisms,
which reflects the invariant symplectic volume form on the underlying quantum space.
Any such frame leads to a rank 3 tensor which is identified with the torsion of the 
Weitzenb\"ock connection. The trace of this tensor is related to the dilaton, and the 
antisymmetric part reduces on-shell to a scalar field identified as axion.
These two scalar fields satisfy second-order equations, sourced by the torsion.

Furthermore, we find a simple spherically invariant static solution for the 
equations of motion, which is localized at some point in space. 
This solution reproduces the linearized Schwarzschild solution, but deviates from 
the full Schwarzschild solution at higher order. 
In particular, 
there is no horizon, and there is no singularity at the origin, more precisely 
the singularity is mild and integrable. The effective energy-momentum tensor 
is smoothly distributed, and can be attributed to the dilaton as well as the 
time component of the frame field. Therefore this solution is not associated  
to matter, but should be viewed as a vacuum  or perhaps as ''remnant`` solution, 
with no counterpart in general relativity. Hence this gravity 
theory is richer than general relativity, which is potentially very interesting.

The extra degrees of freedom of the present theory comprise not only the dilaton and axion fields, but also a tower of higher-spin fields, reminiscent of Vasiliev theory. 
This leads to many open questions.
The equations derived for the frame should be viewed as equations for 
higher-spin valued frames, and it is not evident if the higher-spin 
contributions to the frame can always be eliminated to reproduce some given metric. 
For the present solution, we show that this is indeed the case.
However, in general the  problem of reconstructing frames 
and the role of the higher-spin fields remains to be clarified.

It is tempting to speculate about the feasibility of the present framework as a physical 
theory. From this point of view,
the deviation from Ricci-flatness is perhaps the most interesting 
-- and challenging -- feature:
On the one hand it opens up the possibility for geometric explanations of 
the open problems such as dark energy and dark matter. On the other hand, it 
remains to be seen if the theory can meet the precision tests of gravity.
In particular, the  analog of the full Schwarzschild solution arising from
matter at the origin is still to be found. It may also be that a
proper description of gravity in the presence of matter requires
the inclusion of quantum effects, such as an induced Einstein-Hilbert action 
as discussed previously 
\cite{Sperling:2019xar,Steinacker:2010rh,Sakharov:1967pk}. 
In either case, the tools developed in 
this paper should provide a useful basis for further work towards 
a more complete understanding of gravity in this remarkable model.

\paragraph{Acknowledgments.}

We would especially like to thank Sergio H\"ortner for collaboration on this project.
We would  also like to thank Yuhma Asano for useful discussions and collaboration, and
for pointing out a sign error in a previous version.
The work of HS was supported by the Austrian Science Fund (FWF) grant P32086. 

\appendix

\section{Conventions and useful formulas}

\paragraph{Levi-Civita symbol.} The Levi-Civita symbols will be used with the following convention
\begin{align}
 \varepsilon_{0123} = 1 = -\varepsilon^{0123}
 \label{epsilon-id}
\end{align}
so that 
\begin{align}
 \varepsilon_{\nu\r\s\m} G^{\nu\nu'} G^{\s\s'}G^{\m\m'} G^{\r\r'} = |\det G^{\mu\nu}| \varepsilon^{\nu\r'\s\m'} \ .
  \label{epsilon-lift}
\end{align}

\paragraph{Contraction of contorsion.}
We will also need the following contraction formulas 
which were obtained in  (7.54) and  (7.55) of \cite{Steinacker:2020xph}  
\begin{align}
  \tensor{K}{^\s^\r^\mu}  \tensor{K}{_\r_\s_\nu}  
 &= \frac 14 \Big(2\tensor{T}{^\mu^\s^\r} (\tensor{T}{_\nu_\r_\s} + \tensor{T}{_\nu_\s_\r})
  - \tensor{T}{^\r^\s^\mu}\tensor{T}{_\r_\s_\nu} \Big)
  \label{KK-contract} \\
 \tensor{K}{_\s^\r^\mu}  \tensor{K}{_\r^\s_\mu}  
    &=  \frac 14  \tensor{T}{^\mu^\s^\r} (2\tensor{T}{_\mu_\r_\s} + \tensor{T}{_\mu_\s_\r}) \ .
    \label{KK-contract-full}
 \end{align}

\paragraph{Dilaton and densities.} We  elaborate the relation between the dilaton $\rho$
(which is a scalar function) and the metric and symplectic densities.
To this end, recall that  $\r$ is defined by 
$\r^4 = |\g|^{-1}\r_M^2 = \r^8 |G|^{-1}\r_M^2$, hence
\begin{align}
 \r^{-2}   &= \sqrt{|G|}^{-1}\r_M \ .
\end{align}
Here $\r_M$ is the $SO(4,1)$-invariant volume (density) on $H^4$, 
which  is related to the  dilaton $\bar\r$ of the cosmic background via the analogous relation
\begin{align}
   (\bar\r)^{-2}   &= \sqrt{|\bar G|}^{-1}\r_M  = \sinh^{-3}(\eta)
\label{rho-M}
\end{align}
cf.~(7.23) in \cite{Steinacker:2020xph}.
Taking the ratio gives 
\begin{align}
 \frac{\bar\r^2}{\r^2} &= \frac{\sqrt{|\bar G|}}{\sqrt{|G|}}
\end{align}
hence
\begin{align}
 \sqrt{|G|}\r^{-2} = \sqrt{|\bar G|} \bar\r^{-2} = 
 \r_M  \ ( = \sinh^{-1}(\eta))
 \label{rho-M-G-relation}
\end{align}
where the last equality holds only in Cartesian coordinates.

%
%

\section{Divergence constraint and antisymmetric torsion}

\subsection{Divergence constraint}
\label{sec:div}

We start with the following identity 
\begin{align}
\tensor{T}{_{\mu}_{\nu}^\nu} = 
\tensor{T}{_{\mu}_{\nu}_{\dot\a}} \tensor{E}{^{\dot\a}^{\nu}} 
  &= (\del_\mu \tensor{E}{^{\dot\a}_{\nu}} - \del_\nu \tensor{E}{^{\dot\a}_{\mu}})  \tensor{E}{_{\dot\a}^{\nu}} \nn\\ 
&= \tensor{\cE}{_{\dot\a}^{\nu}}\del_\mu \tensor{\cE}{^{\dot\a}_{\nu}} 
  + \tensor{\cE}{^{\dot\a}_{\mu}}\del_\nu \tensor{\cE}{_{\dot\a}^{\nu}}
  - \frac{3}{\r}\del_\mu\r
  \label{del-E-id-1}
\end{align}
where $\cE$ is the effective frame \eq{eff-frame}.
Together with the relation $-\frac{2}{\r}\del_\mu\r 
 = \tensor{T}{_{\mu}_{\nu}^\nu}$  \eq{tilde-T-T-contract}
for the frame in the matrix model (in the asymptotic regime), we obtain
  \begin{align}
  -\frac{2}{\r}\del_\mu\r 
  &= \tensor{\cE}{_{\dot\a}^{\nu}}\del_\mu \tensor{\cE}{^{\dot\a}_{\nu}} 
  + \tensor{\cE}{^{\dot\a}_{\mu}}\del_\nu \tensor{\cE}{_{\dot\a}^{\nu}}
  - \frac{3}{\r}\del_\mu\r \ . 
  \label{del-E-id-2}
 \end{align}
The first term on the rhs can be rewritten as
\begin{align}
\del_\mu\ln(\sqrt{|G|}) = 
 \det(\tensor{\cE}{^{\dot\a}_{\nu}})^{-1}\del_\mu\det(\tensor{\cE}{^{\dot\a}_{\nu}})
  &=  \tensor{\cE}{_{\dot\a}^{\nu}}\del_\mu \tensor{\cE}{^{\dot\a}_{\nu}} 
   \label{del-E-id-3}
\end{align}
so that 
\begin{align}
  0 &= \del_\mu \ln\sqrt{|G|} + \tensor{\cE}{^{\dot\a}_{\mu}}\del_\nu \tensor{\cE}{_{\dot\a}^{\nu}} - \frac{1}{\r}\del_\mu\r \nn\\
 &= \frac{1}{\sqrt{|G|}}\del_\mu\sqrt{|G|} 
 + \r^{2}\tensor{E}{^{\dot\a}_{\mu}}\del_\nu(\r^{-2} \tensor{E}{_{\dot\a}^{\nu}}) \nn\\
  &=  \frac{\r^{2}}{\sqrt{|G|}}\tensor{E}{^{\dot\a}_{\mu}}\del_\nu \big(\sqrt{|G|} \r^{-2}\tensor{E}{_{\dot\a}^{\nu}}\big)
  \label{del-E-id-4}
\end{align}
which\footnote{This was already shown in (7.18) in \cite{Steinacker:2020xph}.}
results in the divergence constraint~\eq{frame-div-free},
\begin{align}
\boxed{ \ 
  \del_\nu \big(\sqrt{|G|} \r^{-2}\tensor{E}{_{\dot\a}^{\nu}}\big) 
  = 0 =  \nabla^{(G)}_\nu(\r^{-2}\tensor{E}{_{\dot\a}^{\nu}})
  \ .
 }
 \label{frame-div-free-app}
\end{align} 
This means that the $\r^{-2}\tensor{E}{_{\dot \a}^\mu}$ are 
volume-preserving vector fields.
It can be viewed as a gauge-fixing relation analogous to a
Lorentz gauge condition, which reflects the fact that there is no manifest 
local Lorentz invariance in the present framework.

The divergence constraint can also be formulated in terms of differential forms. Because of $\tensor{E}{_{\dot{\alpha}}^{\nu}}=\rho^{2}G^{\nu \mu}E_{\dot{\alpha}\mu}$, we find, using the Hodge star operation with respect to $G$,
\begin{equation}
d ( * E_{\dot{\alpha}}) = 0\ .
\end{equation}

\paragraph{Alternative derivation.}

The origin of the divergence constraint \eq{frame-div-free} from the 
Jacobi identity can be seen as follows: 
\begin{align}
 \tensor{E}{_{\dot \a}^\mu} &\sim -\theta^{\mu\nu}\del_\nu Z_{\dot \a}  \nn\\
 \del_\mu(\r_M\tensor{E}{_{\dot \a}^\mu}) &\sim - \r_M\theta^{\mu\nu}\del_\mu\del_\nu Z_{\dot \a} =  0 
\end{align}
using $\del_\mu(\r_M\theta^{\mu\nu}) = 0$, which holds in the asymptotic regime. 
Together with \eq{rho-M-G-relation}, we recover
\begin{align}
 \del_\mu\big(\sqrt{|G|}\r^{-2}\tensor{E}{_{\dot \a}^\mu}\big) = 0 \ .
\end{align}
Yet another derivation is obtained using (6.24) in \cite{Steinacker:2019fcb}, which implies that 
 $\tensor{E}{_{\dot \a}^\mu} = \{Z_{\dot \a},x^\mu\}_0$ for $Z_{\dot \a} \in\cC^1$
satisfies
\begin{align}
0 &=  \bar\nabla_\nu(\sinh^{-3}(\eta) \tensor{E}{_{\dot \a}^\mu} ) 
  = \frac{1}{\sqrt{|\bar G|}}\del_\nu\Big(\sinh^{-3}(\eta)\sqrt{|\bar G|} \tensor{E}{_{\dot \a}^\nu}\Big) \nn\\
  &= \frac{1}{\sqrt{|\bar G|}}\del_\nu\Big(\sinh^{-1}(\eta)\tensor{E}{_{\dot \a}^\nu}\Big) \
  = \frac{1}{\sqrt{|\bar G|}}\del_\nu\Big(\sqrt{|G|}\r^{-2} \tensor{E}{_{\dot \a}^\nu}\Big) 
\end{align} 
where $\bar G$ is the effective metric of the  FLRW solution $\cM^{3,1}$,
and $x^\mu$ are Cartesian coordinates.

\paragraph{Divergence constraint and diffeomorphisms.}

As a  consistency check, we verify that 
the divergence constraint is preserved by the diffeomorphisms which arise 
from gauge transformations $\{\L,.\}$ in the matrix model.
Indeed,
the frame transforms under a diffeomorphism generated by the vector field 
$\xi^\mu = \{\L,x^\mu\}$ as 
\begin{align}
 \d_\L \tensor{E}{_{\dot\a}^{\mu}} = \cL_\xi \tensor{E}{_{\dot\a}^{\mu}}
 &= \xi^\nu \del_\nu \tensor{E}{_{\dot\a}^{\mu}} 
 -  \tensor{E}{_{\dot\a}^{\nu}} \del_\nu \xi^\mu 
 \end{align}
 so that the divergence constraint transforms as
 \begin{align}
 \d_\L  \del_\mu \big(\sqrt{|G|} \r^{-2}\tensor{E}{_{\dot\a}^{\mu}}\big) 
 &=  \del_\mu \big(\sqrt{|G|} \r^{-2}\xi^\nu \del_\nu \tensor{E}{_{\dot\a}^{\mu}} \big)
 -  \del_\mu \big(\sqrt{|G|} \r^{-2} \tensor{E}{_{\dot\a}^{\nu}} \del_\nu \xi^\mu \big) \nn\\
 &=  \del_\mu \del_\nu\big(\sqrt{|G|} \r^{-2}\xi^\nu  \tensor{E}{_{\dot\a}^{\mu}} \big)
 -  \del_\mu  \del_\nu\big(\sqrt{|G|} \r^{-2} \tensor{E}{_{\dot\a}^{\nu}} \xi^\mu \big) \nn \\
 & \quad - \del_\mu \big(\del_\nu\big(\sqrt{|G|} \r^{-2}\xi^\nu\big)  \tensor{E}{_{\dot\a}^{\mu}} \big) \nn\\
 &= - \del_\mu \big(\del_\nu\big(\sqrt{|G|} \r^{-2}\xi^\nu\big)  \tensor{E}{_{\dot\a}^{\mu}} \big) \ 
\end{align}
using \eq{frame-div-free-app}.
This reduces precisely to the constraint for the diffeomorphisms (7.3) in \cite{Steinacker:2020xph}
\begin{align}
 \sqrt{|G|} \nabla^{(G)}_\nu \big(\r^{-2}\xi^\nu\big)
  &=\del_\nu\big(\sqrt{|G|} \r^{-2}\xi^\nu\big)
 = \del_\nu\big(\r_M\xi^\nu\big)
 = \del_\nu\big(\sqrt{|\bar G|} \bar\r^{-2}\xi^\nu\big) \nn\\
 &= \sqrt{|\bar G|} \nabla^{(\bar G)}_\nu \big(\bar\r^{-2}\xi^\nu\big)
  = \sqrt{|\bar G|} \nabla^{(\bar G)}_\nu \big(\b^3\xi^\nu\big) = 0
  \label{diffeo-constraint}
\end{align}
noting that $\sqrt{|G|}\r^{-2} = \r_M $, 
where  $\bar G$ is the cosmic background metric and
$\b^3=\sinh^{-3}\eta = \bar\r^{-2}$.
Therefore the divergence constraint is consistent  with  the  
volume-preserving diffeos arising in the model.
One can also check that it is compatible with 
the e.o.m.\ for the frame  \eq{eom-torsion-noAS}.

\subsection{The totally antisymmetric torsion $T^{(AS)}$}
\label{sec:AS-torsion-app}

The totally antisymmetric torsion is defined by \eq{T-AS}
\begin{align}
  \tensor{T}{^{(AS)}^\nu_\r_\mu}  &=  \tensor{T}{^\nu_\r_\mu} + \tensor{T}{_\mu^\nu_\r} + \tensor{T}{_\r_\mu^\nu} \ .
\end{align}
Then $G_{\nu\nu'} \tensor{T}{^{(AS)}^{\nu'}_{\r\mu}}$ is totally antisymmetric in the indices $\nu ,\rho ,\mu$ and can naturally be interpreted as a 3-form $T^{(AS)}$.
It is related by the Hodge star $\star$ corresponding to $G_{\mu\nu}$ to a 1-form $T_{\s} dx^\s$
via
\begin{align}
 T^{(AS)} := \frac 16 G_{\nu \nu'}\tensor{T}{^{(AS)}^{\nu'}_\r_\mu}dx^\nu \wedge dx^\r \wedge dx^\mu = \star (T_{\s} dx^\s) \ .
\label{TAS-def}
\end{align}
In coordinates, this amounts to 
\begin{align}
 \tensor{T}{^{(AS)}^\nu_\r_\mu} &= -\sqrt{|G|}G^{\n\n'}\varepsilon_{\n'\r\mu\s} G^{\s\s'} T_{\s'},  \nn\\
 \varepsilon^{\n'\r\mu\k}G_{\n'\n}\tensor{T}{^{(AS)}^\nu_\r_\mu} 
  &= 6\sqrt{|G|} G^{\k\s'}T_{\s'}
 \label{T-AS-dual}
\end{align}
using the conventions \eq{epsilon-id}.
This implies
\begin{align}
 \tensor{T}{^{(AS)}^\s_\r_\mu} \tensor{T}{^{(AS)}_\s^\r_\nu}
  &= |G|G^{\s\s'}\varepsilon_{\s'\r\mu\k}  G^{\r\r'} 
  \tensor{\varepsilon}{_\s_{\r'}_\nu_\eta}  G^{\k\k'}T_{\k'} G^{\eta\eta'}T_{\eta'} \nn\\
  &= \varepsilon^{\s\r\mu'\k}  
  \tensor{\varepsilon}{_\s_{\r}_\nu_\eta} G_{\mu'\mu} T_{\k} G^{\eta\eta'}T_{\eta'} \nn\\
  &= 2 (T_{\nu} T_{\mu} - G_{\nu\mu} G^{\k\eta}T_{\k} T_{\eta} ) 
  \label{TAS-partial-contract}
\end{align}
using \eq{epsilon-lift}, and the contraction gives
\begin{align}
\tensor{T}{^{(AS)}^\s_\r_\mu} \tensor{T}{^{(AS)}_\s^\r_\nu} G^{\mu\nu}
  &= - 6 T_{\nu} T_{\mu}  G^{\mu\nu} \ .
 \label{TAS-contract-1}
\end{align}
We also note the following  identity
\begin{align}
 \tensor{T}{^{(AS)}^\s_\r_\mu} \tensor{T}{^{(AS)}_\s^\r_\nu}
%
%
&= -2( \tensor{T}{_\mu^\s_\r} \tensor{T}{^\r_\s_\nu}
   + \tensor{T}{_\nu^\s_\r} \tensor{T}{^\r_\s_\mu})
   + 2 (\tensor{T}{_\mu^\s_\r}\tensor{T}{_\nu_\s^\r}
    - \tensor{T}{_\mu^\s_\r}\tensor{T}{_\nu^\r_\s}) 
    +  \tensor{T}{^\s_\r_\mu}\tensor{T}{_\s^\r_\nu}  \ .
\label{TAS-contract-id1}
\end{align}
Contracting this with $G^{\mu\nu}$ gives 
\begin{align}
  \tensor{T}{^{(AS)}^\s_\r_\mu} \tensor{T}{^{(AS)}_\s^\r_\nu}G^{\mu\nu}
  &=  3\Big(\tensor{T}{_\mu^\s_\r}\tensor{T}{_\nu_\s^\r} G^{\mu\nu}
  - 2\tensor{T}{^\s_\mu_\r} \tensor{T}{_\s^\r_\nu} G^{\mu\nu}\Big) \nn\\
 \tensor{T}{^\s_\mu_\r} \tensor{T}{_\s^\r_\nu} G^{\mu\nu}
  &= \frac 12\tensor{T}{_\mu^\s_\r}\tensor{T}{_\nu_\s^\r} G^{\mu\nu}
  - \frac 16 \tensor{T}{^{(AS)}^\s_\r_\mu} \tensor{T}{^{(AS)}_\s^\r_\nu}G^{\mu\nu} \nn\\
   &= \frac 12\tensor{T}{_\mu^\s_\r}\tensor{T}{_\nu_\s^\r} G^{\mu\nu}
  + T_{\nu} T_{\mu}  G^{\mu\nu} \ .
  \label{TAS-contract}
\end{align}
We will also need the  formulas
\begin{align}
\tensor{T}{_\mu_\s^\nu}\tensor{T}{_\nu^\s_\r} 
- \tensor{T}{_\r_\s^\nu}\tensor{T}{_\nu^\s_\mu}
 &= -\frac 12(\tensor{T}{^{(AS)}^{\nu}_{\s}_{\mu}} \tensor{T}{_\nu^\s_\r}  - (\mu\leftrightarrow\r))
 \label{TT-id-2}
\end{align}
and 
\begin{align}
 2 \tensor{K}{_\nu_\r^\s}\tensor{T}{_\s_\mu^\nu}
  +2 \tensor{K}{_\nu_\mu^\s}\tensor{T}{_\r_\s^\nu}
   %
 &= -\frac 12(\tensor{T}{^{(AS)}^{\nu}_{\s}_{\mu}} \tensor{T}{_\nu^\s_\r}  - (\mu\leftrightarrow\r)) \ .
 \label{TT-id-3}
\end{align}

\paragraph{The divergence of $T_\mu$.}

The divergence of $T_{\s}$ satisfies the identity
\begin{align}
&  G^{\k\s}\nabla^{(G)}_\k T_{\s}\nn \\
&\qquad = \frac 16  \sqrt{|G|}^{-1} \varepsilon^{\n'\r\mu\k}G_{\n'\n} 
 \nabla^{(G)}_\k \tensor{T}{^{(AS)}^\nu_\r_\mu} \nn\\
 &\qquad = \frac 12 \sqrt{|G|}^{-1} \varepsilon^{\n\r\mu'\k}G_{\m'\m} 
 \nabla^{(G)}_\k \tensor{T}{_\nu_\r^\mu} \nn\\
 &\qquad = \frac 12 \sqrt{|G|}^{-1} \varepsilon^{\n\r\mu'\k}G_{\m'\m} 
 \nabla^{(G)}_\k \Big((\nabla^{(G)}_\nu \tensor{E}{^{\dot\a}_\r} - \nabla^{(G)}_\r \tensor{E}{^{\dot\a}_\nu})
  \tensor{E}{_{\dot\a}^\mu}\Big) \nn\\
  &\qquad = \frac 12 \sqrt{|G|}^{-1} \varepsilon^{\n\r\mu'\k}G_{\m'\m} 
 \Big(\frac 12 \tensor{\cR}{_\k_\n^\s_\r }\tensor{E}{^{\dot\a}_\s}
 - \frac 12 \tensor{\cR}{_\k_\r^\s_\n }\tensor{E}{^{\dot\a}_\s}
  + (\nabla^{(G)}_\nu \tensor{E}{^{\dot\a}_\r} - \nabla^{(G)}_\r \tensor{E}{^{\dot\a}_\nu})
  \nabla^{(G)}_\k \tensor{E}{_{\dot\a}^\mu}\Big) \nn\\
  &\qquad = \frac 12 \sqrt{|G|}^{-1} \varepsilon^{\n\r\mu\k}
 \Big( (\nabla^{(G)}_\nu \tensor{E}{^{\dot\a}_\r} - \nabla^{(G)}_\r \tensor{E}{^{\dot\a}_\nu})
  \nabla^{(G)}_\k (\r^2\tensor{E}{_{\dot\a}_\mu})\Big)  \nn\\
  &\qquad = \frac 14 \r^2\sqrt{|G|}^{-1} \varepsilon^{\n\r\mu\k}
 \tensor{T}{_\nu_\r^{\dot\a}} \tensor{T}{_\k_\mu_{\dot\a}}
  +  \frac 12 (\del_\k \r^2)\sqrt{|G|}^{-1} \varepsilon^{\n\r\mu\k}
 \tensor{T}{_\nu_\r_\mu} \nn\\
  &\qquad = \frac 14 \r^2\sqrt{|G|}^{-1} \varepsilon^{\n\r\mu\k}
 \tensor{T}{_\nu_\r^{\dot\a}} \tensor{T}{_\k_\mu_{\dot\a}}
  +  \frac 16 \r^{-2}\del_\k \r^2\sqrt{|G|}^{-1} \varepsilon^{\n\r\mu\k}
   G_{\mu\mu'} \tensor{T}{^{(AS)}_\nu_\r^{\mu'}}\nn\\
  &\qquad = \frac 14 \r^2\sqrt{|G|}^{-1} \varepsilon^{\n\r\mu\k}
 \tensor{T}{_\nu_\r^{\dot\a}} \tensor{T}{_\k_\mu_{\dot\a}}
  + 2 \r^{-1} \del_\k \r   G^{\k\k'} T_{\k'}
  \label{div-Tmu-id}
\end{align}
which follows from \eq{T-AS-dual}
using the algebraic Bianchi identity for the Riemann tensor.
Multiplied by $\r^{-2}$ this leads to 
the identity
\begin{align}
\tilde\r\,\varepsilon^{\n\r\mu\k}
 \tensor{T}{_\nu_\r^{\dot\a}} \tensor{T}{_\k_\mu_{\dot\a}}
  &= 4\sqrt{|G|}\, \tilde\r\, G^{\mu\nu}\nabla^{(G)}_\mu(\r^{-2} T_{\nu}) \ ,
\end{align}
and using the e.o.m.~\eq{T-del-onshell} for $T_\mu$ this implies  
\begin{align}
 \tilde\r\,\varepsilon^{\n\r\mu\k}
 \tensor{T}{_\nu_\r^{\dot\a}} \tensor{T}{_\k_\mu_{\dot\a}}
  &= 4\sqrt{|G|}\, \tilde\r\, G^{\mu\nu}\nabla^{(G)}_\mu(\r^{-4} \del_{\nu}\tilde\r) \ .
\end{align}
As a by-product, we obtain the following interesting identity 
\begin{align}
 S_{\tilde E} &= -\int d^{4}x \tilde\r\,\varepsilon^{\n\r\mu\k}
 \tensor{T}{_\nu_\r^{\dot\a}} \tensor{T}{_\k_\mu_{\dot\a}}
  = 4\int d^{4}x \sqrt{|G|}\, \r^{-4} G^{\mu\nu}\del_\mu\tilde\r \del_{\nu}\tilde\r
  = 4 S_{\tilde\r} \ .
\end{align}
Since the first term has the structure $\int \tilde\r\, dE \wedge dE$,
this suggests to view $\tilde\r$ as an axion associated to the frame.

\section{6-dimensional configuration space and constraints}
\label{sec:6d-config}

\subsection{General setup}

We want to describe the 6-dimensional symplectic space $\cM^{(6)}$ 
\eq{BG-bundle} underlying the cosmic background solution
more explicitly. It is described through the 8 functions 
$x^\mu$ and $t^\mu$ subject to the constraints \eq{xt-constraints},
which transform covariantly under the global $SO(3,1)$ isometry group
of the $k=-1$ FLRW  spacetime $\cM^{3,1}$.
To exhibit the $S^2$ bundle structure over $\cM^{3,1}$,
it is useful to  consider $\cM^{(6)} $ as Cartesian product\footnote{This description 
misses the double cover structure of $\cM^{(6)}\cong \C P^{1,2}$ \cite{Sperling:2019xar}, but it captures the structure 
after the Big Bounce.}
\begin{align}
 \cM^{(6)} \cong \R^3_x \times \R^3_p
\end{align}
described by the space-like 
Cartesian coordinates $x^i, p^j$, while $x^0$ and $p^0$ are determined by the 
constraints.  This is most transparent in terms of the re-scaled generators 
\begin{align}
 p^\a = \tilde R R t^\a
 \label{tilde-t-def}
\end{align}
with Poisson brackets
\begin{align}
 \{p^\mu,x^\nu\} = \tilde R x^4\eta^{\mu\nu} \ ,\qquad  x^4 = \sqrt{x_\mu x^\mu + R^2} \ .
\end{align}
Then the constraints  \eq{xt-constraints} take the form
\begin{align}
 p_0 x^0  &= p_k x^k \  , \nn\\
-(p_0)^2 + p_i p_i  &=  (x_0)^2 - r^2 
\label{constraints-6D}
\end{align}
(sum over $i$ is understood)
where 
\begin{align}
 r^2 = x_i x_i \ .
\end{align}
The constraints can be written as 
\begin{align}
 (p_0 + x_0)^2 &= p_i p_i + r^2 + 2 p_k x^k = |\vec x+\vec p|^2 \nn\\
 (p_0 - x_0)^2 &= p_i p_i + r^2 - 2 p_k x^k = |\vec x-\vec p|^2
\end{align}
which is solved by 
\begin{align}
 x_0 &= \frac 12(|\vec x+\vec p| + |\vec x-\vec p|) \ > 0 \nn\\
 p_0 &= \frac 12(|\vec x+\vec p| - |\vec x-\vec p|) \ 
 \label{x0-p0-formulas}
\end{align}
which reproduces the cosmic background (after the Big Bounce). 
Note that $p_0$ can take either sign, consistent with \eq{constraints-6D}, and
the triangle inequality applied to \eq{x0-p0-formulas} implies the bound 
\begin{align}
 |p_0| \leq |\vec x| \ .
 \label{p0-bounds}
\end{align}
For a reference point $\xi\in\cM^{3,1}$ with $\vec x=0$, this gives 
\begin{align}
 x_0 = |p|, \quad p_0 = 0 \ .
\end{align}
 More generally in a local region near $\xi$ with
 $|p| = x_0 \gg r$, we have  
\begin{align}
 x^0 &= \frac 12 \big(\sqrt{p^2+r^2+2px} + \sqrt{p^2+r^2-2px}\big) 
  \approx|p| \nn\\
 p_0 &= \frac 12 \big(\sqrt{p^2+r^2+2p x} - \sqrt{p^2+r^2-2px}\big) 
  \approx \frac{p x}{|p|} \approx \frac{px}{x_0} 
  \label{scalar-quant-relations}
\end{align}
up to corrections suppressed by $\frac{r}{x^0}$.
Therefore $|x^0| \approx |p| \approx const$ is essentially the cosmic time, which completes the 
spacetime coordinates on $\cM^{3,1}$
\begin{align}
 x^\mu=(x^0\approx |p|, x^i) \ ,
\end{align}
while the $S^2$  fiber is described by the two transversal $p^i$.

\subsection{$SO(3)$-invariant functions and Poisson brackets}
\label{sec:so3-inv-structure}

Consider the subalgebra of $SO(3)$-invariant functions of $r^2, x^0, p^0$
using the notation \eq{tilde-t-def}.
Here $r^2, x^0$ are viewed as elements of the algebra
$\cC^0$ describing the spin 0 sector of functions on $\cM^{3,1}$, 
while $p^0 \in \cC^1$ is a higher spin generator
 (since it transforms non-trivially under  the local $SO(3)$).
The Poisson brackets are 
\begin{align}
 \{r^2,p^0\} &= 0   \nn\\
 \{r^2,x^0\} 
  &= - 2 \tilde R x^4 p^0  \nn\\
 \{p^0,x^0\} &= - \tilde R x^4 \ .
 \label{invar-brackets}
\end{align}
They span a space of functions in 3 variables. Therefore there must be
a central generator, and the symplectic leaves are 2-dimensional. 
This  generator is  found to be
\begin{align}
\boxed{
 \{\chi,.\} = 0\ , \qquad \chi := r^2 - (p^0)^2 = x_0^2 - |\vec p|^2
 }
 \label{chi-def}
\end{align}
which is central within the $SO(3)$-invariant subalgebra.
Indeed,
\begin{align}
 \{\chi,r^2\} &= - 2p^0\{p^0,r^2\} = 0   \nn\\
 \{\chi,x^0\} &= \{r^2,x^0\} - 2p^0\{p^0,x^0\}  \nn\\
  &= -2\tilde R x^4 p^0 + 2\tilde R x^4p^0 = 0 \ .
\end{align}
We also note that \eq{p0-bounds}
implies 
\begin{align}
 r^2 \geq \chi \geq 0 \ 
\label{chi-pos}
\end{align}
and in particular
\begin{align}
  \qquad p^0 = \pm \sqrt{r^2 - \chi} \ .
  \label{p0-r-chi}
\end{align}

\section{Geometric energy-momentum tensor}
\label{sec:app-e-m-tensor}

It was shown in (eq.~(5.70) in \cite{Steinacker:2020xph})  that 
the Ricci tensor in vacuum satisfies the following vacuum equation of motion 
\begin{align}
  \cR_{\nu\mu} &= - \frac 12 (\tensor{T}{_\r^{\d}_\mu}\tensor{T}{_\nu_\d^\r} 
  + \tensor{T}{_\r^{\d}_\nu}\tensor{T}{_\mu_\d^\r})
  - \tensor{K}{_\d^\r_\nu}\tensor{K}{_\r^\d_\mu}
  + 2\r^{-2}\del_\nu\r \del_\mu\r \nn\\
&\quad  + G_{\nu\mu} \big(\r^{-2} m^2 - \frac 12 \tensor{T}{_\nu^{\s}_\d}\tensor{T}{_\s_\r^\nu} G^{\d\r} \big) \ .
 \label{Ricci-vacuum-2}
\end{align}
The first three terms can be rewritten using \eq{KK-contract} and \eq{TAS-contract-id1} 
as  
\begin{align}
 & - \frac 12\tensor{T}{_\mu_\s^\r} \tensor{T}{_\r^\s_\nu} 
  - \frac 12\tensor{T}{_\nu_\s^\r} \tensor{T}{_\r^\s_\mu}
  -\tensor{K}{_\s^\r_\mu}  \tensor{K}{_\r^\s_\nu}  
%
  = \frac 14 \tensor{T}{^{(AS)}^\s_\r_\mu} \tensor{T}{^{(AS)}_\s^\r_\nu}
   - \tensor{T}{_\mu^\s^\r}\tensor{T}{_\nu_\s_\r} \ .
\end{align}
Therefore the on-shell Ricci tensor \eq{Ricci-vacuum-2} can be written  as 
\begin{align}
 \cR_{\nu\mu}
  %
  &= \frac 14 \tensor{T}{^{(AS)}^\s_\r_\mu} \tensor{T}{^{(AS)}_\s^\r_\nu}
   - \tensor{T}{_\mu_\s^\r}\tensor{T}{_\nu^\s_\r}
  + 2\r^{-2}\del_\nu\r \del_\mu\r \nn\\
 &\quad  + \frac 14  G_{\nu\mu} \big(4\r^{-2} m^2 
 + \tensor{T}{^\s_\nu_\d}\tensor{T}{_\s^\nu_\r} G^{\d\r}
  - \frac{1}{3} \tensor{T}{^{(AS)}^\s_\r_\mu} \tensor{T}{^{(AS)}_\s^\r_\nu}G^{\mu\nu}\big) 
  \label{Ricci-torsion-AS}
\end{align}
and 
\begin{align}
 \cR &= -\frac 1{12} \tensor{T}{^{(AS)}^\s_\r_\mu} \tensor{T}{^{(AS)}_\s^\r_\nu} G^{\mu\nu}
  + 2\r^{-2}\del_\mu\r \del^\mu\r + 4\r^{-2} m^2 \ .
  \label{R-TAS-rho-id}
\end{align}
Using \eq{TAS-contract-1} 
and the on-shell relation \eq{T-del-onshell} for $T_\mu$, 
we obtain the on-shell relation 
\begin{align}
 \cR &= \frac 1{2} \r^{-4}G^{\mu\nu}\del_\mu\tilde\r\del_\nu\tilde\r
  + 2\r^{-2}\del_\mu\r \del^\mu\r + 4\r^{-2} m^2 \ .
\label{R-axion-r-onshell}
\end{align}
This implies that solutions with $\cR=0$ can arise  (in the limit $m^2\to 0$)
only if both $\r$ and $\tilde\r$ are constant (or have light-like derivatives).

\paragraph{Effective energy-momentum tensor.}

We now {\em define} the 
effective energy-momentum (e-m) tensor due to torsion in terms of the Einstein equations
(absorbing a factor $8\pi$ in the e-m tensor), which thus
decomposes into different contributions
\begin{align}
 {\bf T}_{\mu\nu} 
 &:= \cR_{\nu\mu} - \frac 12 G_{\mu\nu} \cR  \nn\\
 &= \frac 14 \tensor{T}{^{(AS)}_\s^\r_\mu} \tensor{T}{^{(AS)}^\s_\r_\nu}
   - \tensor{T}{_\mu_\s^\r}\tensor{T}{_\nu^\s_\r}
  + 2\r^{-2}\del_\nu\r \del_\mu\r \nn\\
 &\quad  + \frac 14  G_{\nu\mu} \big(-4\r^{-2} m^2 
 + \tensor{T}{^\s_\nu_\d}\tensor{T}{_\s^\nu_\r} G^{\d\r}
  - \frac{1}{6} \tensor{T}{^{(AS)}_\s^\r_\mu} \tensor{T}{^{(AS)}^\s_\r_\nu}G^{\mu\nu}
  - 4\r^{-2}\del^\mu\r \del_\mu\r ) \nn\\
  &=  {\bf T}_{\mu\nu}[E^{\dot\a}] 
   + {\bf T}_{\mu\nu}[\r] 
    + {\bf T^{(AS)}}_{\mu\nu}[T] 
       - \r^{-2} m^2 G_{\nu\mu} \ .
    \label{e-m-torsion}
\end{align}
Here the energy-momentum tensor of the dilaton is
\begin{align}
 {\bf T}_{\mu\nu}[\r] &= 2\r^{-2}\Big(\del_\nu\r \del_\mu\r - \frac 12 G_{\nu\mu}\del^\s\r \del_\s\r \Big)  \nn\\
  & = 2\big(\del_\nu\s \del_\mu\s - \frac 12 G_{\nu\mu}\del^\k\s \del_\k\s \big) ,
 \qquad \r=e^{-\s} \ .
 \label{e-m-tensor-rho}
\end{align}
For the contribution of $T^{(AS)}$ we can use \eq{TAS-partial-contract} and \eq{TAS-contract-1}, 
so that
\begin{align} 
 {\bf T^{(AS)}}_{\mu\nu}[T]
 &= \frac 14\big(\tensor{T}{^{(AS)}^\s_\r_\mu} \tensor{T}{^{(AS)}_\s^\r_\nu}
      - \frac 16 G_{\nu\mu} (\tensor{T}{^{(AS)}^\s_\r_\kappa} \tensor{T}{^{(AS)}_\s^\r_\lambda} G^{\kappa\lambda}) \big) \nn\\
 &= \frac 12 \Big(  T_\mu T_\nu - \frac 12 G_{\nu\mu} (T^\r T_\r)\Big)  \ .
\end{align}
Using the e.o.m.~\eq{T-del-onshell} for $T_\mu$,  this reduces to the 
 e-m tensor of the axion,
\begin{align}
  {\bf T^{(AS)}}_{\mu\nu}[T] 
 &= \frac 12 \r^{-4}\Big(\del_\mu\tilde\r \del_\nu\tilde\r 
  - \frac 12 G_{\nu\mu} (G^{\s\s'}\del_\s\tilde\r \del_{\s'}\tilde\r)\Big)
  =:  {\bf T}_{\mu\nu}[\tilde\r] \ .
\end{align}
Finally, the e-m tensor of the frame is
\begin{align}
 {\bf T}_{\mu\nu}[E^{\dot\a}] 
&:= -\tensor{T}{_\mu_\s^\r}\tensor{T}{_\nu^\s_\r} 
 + \frac 14 G_{\nu\mu} \tensor{T}{^\s_\nu_\d}\tensor{T}{_\s^\nu_\r} G^{\d\r} \nn\\
  &= - \r^2\big(F_{\mu\s}[E^{\dot\a}]F_{\n\r}[E_{\dot\a}] G^{\r\s} 
  - \frac 14 G_{\nu\mu}( F_{\r\k}[E^{\dot\a}]F_{\s\k'}[E_{\dot\a}] G^{\r\s} G^{\k\k'})  \big)
\label{e-m-tensor-frame}
\end{align}
where 
\begin{align}
 F_{\mu\nu}[E^{\dot\a}] := \del_\mu \tensor{E}{_{\n}^{\dot\a}} - \del_\n \tensor{E}{_{\mu}^{\dot\a}} = \tensor{T}{_\mu_\nu^{\dot\a}}
\end{align}
is the field strength of the frame vector field. Thus 
 ${\bf T}_{\mu\nu}[E^\a]$ has the structure of 
a {\em negative}  Maxwell-like e-m tensor 
\begin{align}
{\bf T}_{\mu\nu}[A] &= F_{\mu\r} F_{\nu\s}G^{\r\s}  - \frac 14 G_{\mu\nu} (F_{\r\s} F^{\r\s})\ ,
\qquad F = dA
\end{align}
from each frame component $E^{\dot\a}$, with opposite sign for space- and time-like 
components. 
Hence it is possible in principle to obtain  Ricci-flat solutions,
provided the various contributions 
with different signs cancel.

\section{Geometric actions and identities}

\subsection{Einstein-Hilbert term from torsion}

We  start from the identity (5.61) in \cite{Steinacker:2020xph} and obtain the Ricci scalar as
\begin{align}
  \cR[G] 
   &= 2\nabla_{(G)}^\mu \tensor{\cK}{_\nu_\mu^\nu} 
        - \tensor{\cK}{_\mu_\r^\mu} \tensor{\cK}{_\nu_\s^\r}  G^{\nu\s}
        + \tensor{\cK}{_\nu_\r^\mu} \tensor{\cK}{_\mu_\s^\r} G^{\nu\s}    \nn\\        
    &=  \tensor{\cK}{_\nu_\r^\mu} \tensor{\cK}{_\mu_\s^\r} G^{\nu\s}
        - \tensor{\cK}{_\mu_\r^\mu} \tensor{\cK}{_\nu_\s^\r}  G^{\nu\s} 
         - 2\nabla_{(G)}^\mu (\r^{-1} \del_\mu\r)   \nn\\
 &=  \tensor{K}{_\nu_\r^\mu} \tensor{K}{_\mu_\s^\r} G^{\nu\s} 
        + 2 \r^{-2} G^{\mu\nu}\del_\mu\r\del_\nu\r 
       - 2\nabla_{(G)}^\mu (\r^{-1} \del_\mu\r) \nn\\
 &=  - \frac 14\tensor{T}{^\mu_\s_\r} \tensor{T}{_\mu_{\s'}^\r} G^{\s\s'} 
 -  \frac 12\tensor{T}{^\mu_\s_\r} \tensor{T}{_\mu^\r_{\s'}} G^{\s\s'} 
        + 2 \r^{-2} G^{\mu\nu}\del_\mu\r\del_\nu\r 
        - 2\nabla_{(G)}^\mu (\r^{-1} \del_\mu\r) 
        \label{R-offshell-T}
\end{align}
using \eq{Levi-contorsion-full},
the contraction relation \eq{tilde-T-T-contract},  and  \eq{KK-contract-full} in the last step.
This is an identity, no equations of motion are used.
We can rewrite the second term using \eq{TAS-contract}
 as
 \begin{align}
   \cR &=  - \frac 12\tensor{T}{^\mu_\s_\r} \tensor{T}{_\mu_{\s'}^\r} G^{\s\s'}
  - \frac 12 T_{\nu} T_{\mu}  G^{\mu\nu}  
        + 2 \r^{-2} G^{\mu\nu}\del_\mu\r\del_\nu\r 
        - 2\nabla_{(G)}^\mu (\r^{-1} \del_\mu\r) \ .
        \label{R-offshell-T-2}
 \end{align}
This is an identity which applies to any off-shell configuration. 
Thus all action terms quadratic in the
torsion can be written
in terms of the Einstein-Hilbert term and the dilaton. 
Similar identities are used in the teleparallel reformulation of general relativity \cite{aldrovandi2012teleparallel}, but the role of the dilaton and the divergence constraint is specific to the present framework.

\subsection{Variation of the  action $S_T$}

Consider the action term for the totally antisymmetric torsion
\begin{align}
  S_{T} &= \int d^{4}x \sqrt{|G|} G^{\mu\nu}T_\mu T_\nu  \ . 
\end{align}
Here $T_\mu$ is defined in \eq{T-AS-dual}, or explicitly
\begin{align}
  T_{\r} = \frac 12\sqrt{|G|}^{-1}\r^2 G_{\r\k}\varepsilon^{\n\s\mu\k}\tensor{T}{_\nu_\s_\mu} \ .
 \label{T-AS-dual-2}
\end{align}
Its variation can be written as
\begin{align}
 2\d  T_{\r}
   &= \sqrt{|G|}^{-1}\r^2
  \big( G_{\r\k}\varepsilon^{\n\s\mu\k}\d\tensor{T}{_\nu_\s_\mu} 
   + \d G_{\r\k} \varepsilon^{\n\s\mu\k}\tensor{T}{_\nu_\s_\mu}\big)
   + \Big(\r^{-2}\d\r^2 + \frac 12 \d G^{\a\b} G_{\a\b}\Big) 2T_{\r} \ . 
\end{align}
Using
\begin{align}
 \varepsilon^{\n\s\mu\k} \d\tensor{T}{_{\nu}_{\s}_{\mu}} 
  &= \varepsilon^{\n\s\mu\k}\Big( (\del_\nu \tensor{E}{^{\dot\a}_\s} - \del_\s \tensor{E}{^{\dot\a}_\nu})\d\tensor{E}{_{\dot\a}_\mu}
    + (\del_\nu \d\tensor{E}{^{\dot\a}_\s} - \del_\s \d\tensor{E}{^{\dot\a}_\nu})\tensor{E}{_{\dot\a}_\mu}\Big)  \nn\\
  &= \varepsilon^{\n\s\mu\k} \Big(\tensor{T}{_\nu_\s^{\dot\a}}\d\tensor{E}{_{\dot\a}_\mu}
    + 2\del_\nu \d\tensor{E}{^{\dot\a}_\s}\tensor{E}{_{\dot\a}_\mu} \Big)
\end{align}
and partial integration we obtain 
\begin{align}
&\int d^{4}x \,\r^2 T_\k \varepsilon^{\nu\mu\s\k} \d\tensor{T}{_{\nu}_{\s}_{\mu}} \nn \\
&\qquad =  \int d^{4}x 
  \Big(\r^2  T_\k \varepsilon^{\nu\mu\s\k} \tensor{T}{_\nu_\s^{\dot\a}}\d\tensor{E}{_{\dot\a}_\mu}
    - 2 \varepsilon^{\nu\mu\s\k} \del_\nu (\r^2 T_\k)\tensor{E}{_{\dot\a}_\mu}
    \d\tensor{E}{^{\dot\a}_\s}
    - \r^2  T_\k \varepsilon^{\nu\mu\s\k}  \tensor{T}{_\nu_\mu^{\dot\a}}
    \d\tensor{E}{_{\dot\a}_\s}\Big)   \nn\\
  &\qquad =  2\int d^{4}x \,\r^2\d\tensor{E}{_{\dot\a}_\mu}
  \Big( T_\k \varepsilon^{\nu\mu\s\k} \tensor{T}{_\nu_\s^{\dot\a}}
    + \tensor{E}{^{\dot\a}_\s}
   \varepsilon^{\nu\mu\s\k} \r^{-2}\del_\nu( \r^2  T_\k)\Big) \ .
\end{align}
Therefore the variation of $S_{T}$ can be written as 
\begin{align}
  \d S_{T} &= 
 \int d^{4}x \sqrt{|G|} \d G^{\mu\nu} \Big(T_\mu T_\nu - \frac 12 G_{\mu\nu} T\cdot T\Big)
   +  \int d^{4}x  \sqrt{|G|} G^{\mu\nu} T_\mu 2\d T_\nu \nn\\
   &= \int d^{4}x \sqrt{|G|} \d G^{\mu\nu} \Big(T_\mu T_\nu + \frac 12 G_{\mu\nu} T\cdot T\Big) \nn\\
 &\quad  + \int d^{4}x \big(\r^2 T_\k\varepsilon^{\n\s\mu\k}\d\tensor{T}{_\nu_\s_\mu}  + \r^2 T_\eta G^{\eta\r} \d G_{\r\k} \varepsilon^{\n\s\mu\k}\tensor{T}{_\nu_\s_\mu}\big) + 2\int d^{4}x \r^{-2}\d\r^2 T\cdot T
\nn\\
   &= \int d^{4}x \sqrt{|G|} \Big(-\d G^{\mu\nu} (T_\mu T_\nu - \frac 12 G_{\mu\nu} T\cdot T)
   + 2 \r^{-2}\d\r^2 T\cdot T \Big) 
   +\int d^{4}x \r^2 T_\k\varepsilon^{\n\s\mu\k}\d\tensor{T}{_\nu_\s_\mu} \nn\\
   &= \int d^{4}x \sqrt{|G|} \Big(-\d G^{\mu\nu} (T_\mu T_\nu - \frac 12 G_{\mu\nu} T\cdot T)
   + 2 \r^{-2}\d\r^2 T\cdot T \Big) \nn\\
 &\quad - 2\int d^{4}x \r^2\d\tensor{E}{_{\dot\a}_\mu}
  \Big( T_\k \varepsilon^{\nu\mu\s\k} \tensor{T}{_\nu_\s^{\dot\a}}
    + \tensor{E}{^{\dot\a}_\s}
   \varepsilon^{\nu\mu\s\k} \r^{-2}\del_\nu( \r^2  T_\k)\Big) 
\end{align}
where $T\cdot T = G^{\mu\nu} T_\mu T_\nu$. 
This is used in \eq{variations-action-1}.

\bibliographystyle{JHEP}
\bibliography{papers}

\end{document}